\documentclass[journal,twoside]{IEEEtran}

\usepackage{amssymb}
\usepackage[cmex10]{amsmath}
\usepackage{graphics}
\usepackage{latexsym}
\newtheorem{Theorem}{Theorem}
\newtheorem{Proposition}{Proposition}
\newtheorem{Lemma}{Lemma}
\newtheorem{Corollary}{Corollary}

\begin{document}

\title{
Inverse-free Berlekamp--Massey--Sakata Algorithm and Small Decoders for Algebraic-Geometric Codes}

\author{Hajime~Matsui,~\IEEEmembership{Member,~IEEE,}
        and~Seiichi~Mita,~\IEEEmembership{Member,~IEEE}
\thanks{Manuscript received April 10, 2007.
This work was partly supported
by the Academic Frontier Project
for Future Data Storage Materials Research
by the Japanese Ministry of Education, Culture, Sports,
Science and Technology (1999--2008).
The material in this paper was presented in part at Hawaii, IEICE and SITA Joint Conference on Information Theory, Hawaii, May 2005,
and at IEEE International Symposium on Information Theory, Seattle, July 2006.}
\thanks{H. Matsui and S. Mita are with
the Department of Electronics and Information Science,
Toyota Technological Institute,
Hisakata 2--12--1, Tenpaku-ku, Nagoya 468--8511, Japan
(e-mail: hmatsui@toyota-ti.ac.jp; smita@toyota-ti.ac.jp).}
}

\markboth{Submitted to IEEE Transactins on Information Theory, 2007}{Matsui \MakeLowercase{\textit{et al.}}: Inverse-free Berlekamp--Massey--Sakata Algorithm and Small Decoders for Algebraic-Geometric Codes
}

\maketitle

\begin{abstract}
This paper proposes a novel algorithm for finding error-locators of
algebraic-geometric codes that can eliminate the division-calculations of
finite fields from the Berlekamp--Massey--Sakata algorithm.
This inverse-free algorithm provides full performance in correcting a
certain class of errors, generic errors, which includes most errors, and can
decode codes on algebraic curves without the determination of unknown
syndromes.
Moreover, we propose three different kinds of architectures that our algorithm can be applied to, and we represent the control operation of 
shift-registers and switches at each clock-timing with numerical simulations.
We estimate the performance in comparison of the total running time and the 
numbers of multipliers and shift-registers in three architectures with those 
of the conventional ones for codes on algebraic curves.
\end{abstract}


\section{Introduction}

\IEEEPARstart{A}{lgebraic-geometric} (AG) codes, especially codes on algebraic curves, are comprehensive generalization of prevailing Reed--Solomon (RS) codes.
They can be applied to various systems by choosing suitable algebraic curves 
without any extension to huge finite (Galois) fields.
In fast decoding of such codes, Berlekamp--Massey--Sakata (BMS) algorithm 
\cite{finding} is often used for finding the location of errors, and the 
evaluation of error-values is done by using outputs of BMS algorithm with 
O'Sullivan's formula \cite{O'Sullivan error values}.

RS codes have the features of high error-correcting capability and less 
complexity for the implementation of encoder and decoder.
On the other hand, codes on algebraic curves have the issues related to the 
size of decoders as well as the operating speed of decoders.
In particular, we notice that RS-code decoders need no inverse-calculator of 
the finite field (no finite-field inverter).
The extended Euclidean algorithm \cite{Sugiyama-Kasahara-Hirasawa-Namekawa} 
for RS codes has no divisions, and this enables us to operate compactly and quickly in calculating error-locator and error-evaluator polynomials.
One inverse computation requires thirteen multiplications in practical 
GF$(2^{8})$ and needs enormous circuit scale.
Thus, it is strongly expected that the fast inverse-free algorithm for AG 
codes will be established, since division operations are inevitable on the 
original BMS algorithm.
In addition, the decoder that has small circuit-size, such as the 
conventional RS decoder, is considered necessary.

In this paper, we propose an inverse-free BMS algorithm, and give a whole 
proof of its adequacy.
Moreover, we propose three kinds of small-sized architectures that generate 
error-locator polynomials for codes on algebraic curves.
We then explain our architectures with model structures and numerical 
examples, and show the practical operation of proposed architectures in 
terms of the control flow of registers and switches at each clock-timing.
The performance is estimated on the total running time and the numbers of 
multipliers and shift-registers for all architectures.

The divisions in the original BMS algorithm appear at the Berlekamp 
transform \cite{Berlekamp}
\begin{equation}
f_{N+1}:=f_{N}-\left(d_{N}/\delta_{N}\right)g_{N}\label{denominator}
\end{equation}
at each $N$-loop in the algorithm, where $f_{N}$, $g_{N}$, and $d_{N}$ are called minimal polynomial, auxiliary polynomial, and discrepancy at $N$, respectively, $N$ runs over $0\le N\le B$ for sufficiently large $B$, and 
$\delta_{N}$ is equal to a certain previous $d_{N}$.
Then the inverse-free BMS algorithm consists of modified Berlekamp 
transforms of the form
\begin{equation}
f_{N+1}:=e_{N}f_{N}-d_{N}g_{N},\label{clear}
\end{equation}
where $e_{N}$ is equal to a certain previous $d_{N}$ in this expression.
Thus the denominator $\delta_{N}$ in \eqref{denominator} is converted into 
the multiplication of $e_{N}$ in \eqref{clear}.
This version of inverse-free BMS algorithm can be proved in the comparable 
line of the original algorithm.
However, there is a significant obstacle to apply this inverse-free 
algorithm to the decoders for AG codes; we have to mention the existence of 
unknown syndromes, namely, the lack of syndrome values to decode errors 
whose Hamming weights are less than or equal to even the basic 
$\left\lfloor(d_\mathrm{G}-1)/2\right\rfloor$, where $d_\mathrm{G}$ is the 
Goppa (designed) minimum distance.
Feng and Rao's paper \cite{Feng-Rao} originally proposed majority logic 
scheme to determine unknown syndromes in the decoding up to 
$\left\lfloor(d_\mathrm{FR}-1)/2\right\rfloor$, where $d_\mathrm{FR}$ is 
their designed minimum distance $\ge d_\mathrm{G}$.
In the sequel, Sakata \textit{et al.} \cite{Sakata-Jensen-Hoholdt} and 
independently K\"otter \cite{K"otter} modified and applied Feng--Rao's 
method to their decoding algorithm.
If the divisions of the finite field are removed from BMS algorithm, one 
cannot execute the determination of unknown syndromes because of breaking 
the generation of candidate values of unknown syndromes for majority voting.
Unfortunately, the elimination of finite-field divisions seemed to be a 
difficult problem in this regard.
For this reason, no inverse-free algorithm for AG codes has been proposed 
until now.

In this research, we effectively overcome this difficulty. Namely, we decode 
such codes with the only known syndrome values from received code-words.
So far the type and amount of errors that could be corrected if one does not determine unknown syndromes have not been clear;
the well-known fact 
up to $\left\lfloor(d_\mathrm{G}-g-1)/2\right\rfloor$ in Peterson-type 
algorithm \cite{Justesen}, where $g$ is the genus of underlying algebraic 
curve, is \textit{not} available for our case of BMS algorithm.
We confirm that a class of generic errors \cite{ISITA04}\cite{generic} 
(independent errors in \cite{independent}) can be corrected up to 
$\left\lfloor(d_\mathrm{FR}-a)/2\right\rfloor$ only with syndromes from 
received words, where $a$ is the minimal pole order of underlying algebraic 
curve: $a=2$ for elliptic curves over arbitrary finite fields and $a=16$ for 
Hermitian curve over GF$(2^{8})$.
Furthermore, we successfully obtain the approximate ratio $(q-1)/q$ of the 
generic errors to all errors in the application of Gr\"obner-basis theory, 
where $q$ is the number of elements in the finite field.
It means that we can decode most of the errors without majority logic scheme 
and voting.
Thus we can realize not only inverse-free error-locator architectures for AG 
codes but also avoiding complicated procedure and transmission of voting 
data among parts of decoders.
Our method is applicable to all former architectures, and is not a go-back 
to the past but a real solution to construct decoders with feasible 
circuit-scale.

\begin{figure}[!t]
\centering
  \resizebox{8.8cm}{!}{\includegraphics{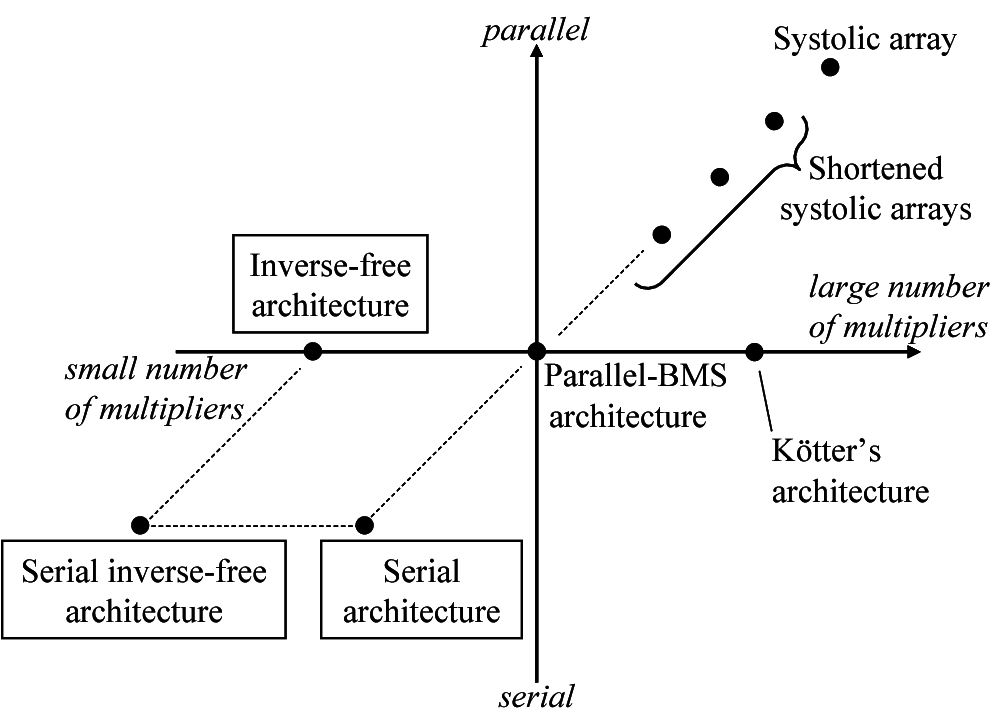}}
\caption{Map of various error-locator architectures implementing BMS (or 
equivalent) algorithm for decoding codes on algebraic 
curves.\label{reduction map}}
\end{figure}

Recently, the BMS algorithm has become more important not only in decoding 
codes on algebraic curves but also in algebraic soft-decision decoding 
\cite{Koetter-Vardy} of RS codes.
Sakata \textit{et al.} \cite{Numakami}\cite{Linz} applied the BMS algorithm to the polynomial interpolation in Sudan and Guruswami--Sudan algorithms 
\cite{Guruswami-Sudan}\cite{Sudan} for RS codes and codes on algebraic 
curves.
Lee and O'Sullivan \cite{Lee-O'Sullivan ISIT06}\cite{perspective} applied 
the Gr\"obner-basis theory of modules, which is related to the BMS 
algorithm, to soft-decision decoding of RS codes.
Our method can be expected to help further structural analysis of these 
methods.

The rest of this paper is organized as follows.
In Section \ref{Preliminaries}, we prepare notations, and define codes on algebraic curves.
In Section \ref{Inverse-free BMS algorithm}, we propose an inverse-free BMS algorithm, and state the main 
theorem for output of the algorithm.
In the next three sections, we describe three types of small-scale 
error-locator architectures, i.e., \textit{inverse-free}, \textit{serial}, and \textit{serial inverse-free architectures;}
the mutual relations among them and past architectures are depicted in 
Fig. \ref{reduction map}.
In Section \ref{Inverse-free architecture}, we describe the inverse-free architecture, and divide it into 
three subsections:
Subsection \ref{Model structure} is an overview, Subsection \ref{Decoding of generic errors} deals with the technique for 
avoiding the determination of unknown syndromes, and Subsection \ref{Simulation} is 
numerical simulation.
In Section \ref{Serial architecture}, we describe the serial architecture using parallel BMS 
algorithm.
In Section \ref{Serial inverse-free architecture}, we describe the serial inverse-free architectures combined 
with the previous methods.
In Section \ref{Performance estimation}, we estimate the total running time and the numbers of 
finite-field calculators for three and past architectures.
Finally, in Section \ref{Conclusions}, we state our conclusions.
In the appendices, we prove the basics of BMS algorithm, the property of 
generic errors, and the main theorem of proposed algorithm.

\begin{figure*}[t]
\centering
  \resizebox{12cm}{!}{\includegraphics{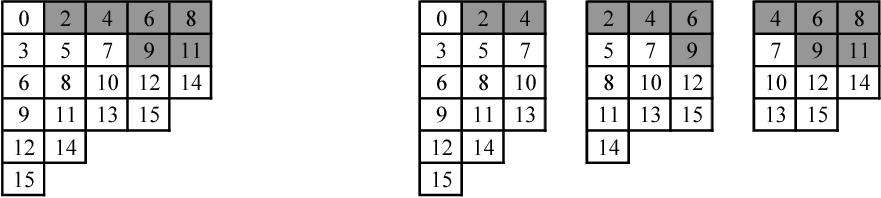}}
\caption{Pole orders on $\Phi(5,15)$ defined by $o(n):=3n_{1}+2n_{2}$, and pole orders on $\Phi^{(0)}(3,15)$, $\Phi^{(1)}(3,15)$, $\Phi^{(2)}(3,15)$. The values in shaded boxes correspond to monomials of the form $x^{n_{1}}y^{n_{2}}$ \textit{not} contained in $L(15P_{(0:0:1)})$ of Klein's quartic curve $x^{3}y+y^{3}+x=0$ over GF($2^3$) (cf. later section \ref{Serial architecture}).\label{Phi(3,15)}}
\end{figure*}

\section{Preliminaries}\label{Preliminaries}

In this paper, we consider one-point algebraic-geometric codes on 
non-singular plane curves over a finite field $K:=\mathbb{F}_{q}$, in 
particular $\mathcal\Omega$-type codes (not $L$-type).
Let ${\mathbb Z}_{0}$ be the set of non-negative integers, and let 
$a,b\in{\mathbb Z}_{0}$ be $0<a\le b$ and $\gcd(a,b)=1$.
We define a C${}_{a}^{b}$ curve $\mathcal{X}$ by an equation
\begin{equation}
D(x,y):=y^{a}+ex^{b}+
\sum_{\substack{(n_{1},n_{2})\in{\mathbb Z}_{0}^{2}\\n_{1}a+n_{2}b<ab}}
\chi_{(n_{1},n_{2})}x^{n_{1}}y^{n_{2}}=0
\label{cabcurve}
\end{equation}
over $K$ with $e\not=0$.
Then the polynomial quotient ring $K[\mathcal{X}]:=K[x,y]/(D(x,y))$ consists 
of all the algebraic functions having no poles except at the unique infinite 
point $P_{\infty}$. Let $\{P_{j}\}_{1\le j\le n}$ be a set of $n$ 
$K$-rational points except $P_{\infty}$. We denote the pole order of $F\in 
K[\mathcal{X}]$ at $P_{\infty}$ as $o(F)$. For $m\in{\mathbb Z}_{0}$, the 
$K$-linear subspace
$$L(mP_{\infty}):=\{F\in K[\mathcal{X}]\mid o(F)\le m\}\cup\{0\}$$
has dimension $m-g+1$, provided $m>2g-2$ by Riemann--Roch theorem, which we 
assume for simplicity in this paper. Our code $\mathcal{C}(m)$ is defined as
$$
\mathcal{C}(m):=
\left\{(c_{j})\in 
K^{n}\left|\,\displaystyle{\sum_{j=1}^{n}}c_{j}F(P_{j})=0,\:\forall\,F\in 
L(mP_{\infty})\right.\right\}.
$$
As shown in \cite{Matsumoto}\cite{Miura}, the class of C${}_{a}^{b}$ curves 
is sufficiently wide and contains almost all well-known plane algebraic 
curves that have many $K$-rational points such as Hermitian codes.
Although Miura in \cite{Miura} defined a more general class ${}_r$C${}_{a}^{b,d}$ including the Klein's quartic curve, we consider mainly C${}_{a}^{b}$ for simplicity.

Throughout this paper, we denote $t$ as the number of correctable errors.
Given a received word $(r_{j})=(c_{j})+(e_{j})$, where $e_{j}\not=0$ 
$\Leftrightarrow$ $j\in\{j_{1},\cdots,j_{t}\}$ corresponding to a set of error-locations $\mathcal{E}=\{P_{j_{\gamma}}\}_{1\le\gamma\le t}$, we need to 
find a Gr\"obner basis \cite{Cox} of the error-locator ideal
$$I(\mathcal{E}):=
\{F\in K[\mathcal{X}]\,|\,F(P_{j_{\gamma}})=0\;\mathrm{for}\;\forall\,
P_{j_{\gamma}}\in\mathcal{E}\}.$$
Then we can obtain $\mathcal{E}$ as the set $\subset\{P_{j}\}_{1\le j\le n}$ 
of common zeros of all the polynomials in the Gr\"obner basis.

For $A\in{\mathbb Z}_{0}$ and $0\le i<a$, let
$$
\Phi^{(i)}(A):=\{n=(n_{1},n_{2})\in{\mathbb Z}_{0}^{2}\,\big|\,
i\le n_{2}<i+A\}
$$
and $\Phi(A):=\Phi^{(0)}(A)$.
Moreover, for $A'\in{\mathbb Z}_{0}$, let
$$\Phi^{(i)}(A,A'):=\{n\in\Phi^{(i)}(A)\,\big|\,o(n)\le A'\}$$
and $\Phi(A,A'):=\Phi^{(0)}(A,A')$.
Fig. \ref{Phi(3,15)} illustrates $\Phi(2a-1,A')$ and $\Phi^{(i)}(a,A')$
for $A'=15$ and $(a,b)=(3,2)$;
although we defined as $a\le b$, it must be generalized into $a>b$ in the 
case of well-known Klein's quartic curve, which is one of the important examples not contained in C${}_{a}^{b}$ curves; we will also take up codes on
this curve later in section \ref{Serial architecture}.
We note that $o(n)\not=o(n')$ if and only if $n\not=n'$ for 
$n,n'\in\Phi^{(i)}(a)$, and this is false for $\Phi(2a-1)$.
Thus $F\in K[\mathcal{X}]$ is uniquely expressed as
\begin{equation}\label{expression}
F(x,y)=\sum_{n\in\Phi(a,o(F))}F_{n}x^{n_{1}}y^{n_{2}}.
\end{equation}
We denote $x^{n_{1}}y^{n_{2}}$ by $z^{n}$ and define 
$o(n):=o(z^n)=n_{1}a+n_{2}b$, where $o(\cdot)$ is defined on both ${\mathbb 
Z}_{0}^{2}$ and $K[\mathcal{X}]$; we remember that 
$o(F)=\max\{o(n)|\,F_{n}\not=0\}$.

From a given received word $(r_{j})$, we calculate syndrome values 
$\{u_{l}\}$ for $l\in\Phi(2a-1,m)$ by 
$u_{l}=\sum_{j=1}^{n}r_{j}z^{l}(P_{j})$, where we have 
$u_{l}=\sum_{\gamma=1}^{t}e_{j_{\gamma}}z^{l}(P_{j_{\gamma}})$ by the 
definition of $\mathcal{C}(m)$.
Our aim is to find $I(\mathcal{E})$ and $(e_{j})$ with $\{u_{l}\}$.

\section{Inverse-free BMS algorithm}
\label{Inverse-free BMS algorithm}
We continue to prepare notations to describe the algorithm.
The standard partial order $\le$ on ${\mathbb Z}_{0}^{2}$ is defined as 
follows: for $n=(n_{1},n_{2})$ and $n'=(n'_{1},n'_{2})\in{\mathbb 
Z}_{0}^{2}$, $n\le n'$ $\Leftrightarrow$ $n_{1}\le n'_{1}$ and $n_{2}\le 
n'_{2}$.
For $l\in\Phi(a,A')$, let $l^{(i)}\in\Phi^{(i)}(a,A')$ be $o(l^{(i)})=o(l)$ 
if there exists such an $l^{(i)}$ for $l$ and $i$. Then $l^{(i)}$ is 
uniquely determined for each $l$ and $i$ if it exists. Note that $l^{(0)}=l$ 
from its definition. Table \ref{l^(i)} illustrates 
$l^{(i)}\in\Phi^{(i)}(3,15)$ for $(a,b)=(3,2)$, where ``$*$'' indicates the 
nonexistence of $l^{(i)}$ from a gap-number in $o(\Phi^{(i)}(a))$.

Before the description of the algorithm, we introduce the important index 
$\overline{\imath}$ for $0\le i<a$ for updating in the algorithm. For $0\le 
i<a$ and $N\in{\mathbb Z}_{0}$, we define a unique integer 
$0\le\overline{\imath}<a$ by $\overline{\imath}\equiv 
b^{-1}N-i\:(\mathrm{mod}\,a)$, where the integer $0\le b^{-1}<a$ is defined 
by $b\,b^{-1}\equiv1\:(\mathrm{mod}\,a)$. If there is 
$l^{(i)}=(l_{1}^{(i)},l_{2}^{(i)})\in\Phi^{(i)}(a)$ with $N=o(l^{(i)})$, 
then $\overline{\imath}=l_{2}^{(i)}-i$ since $l_{2}^{(i)}\equiv 
b^{-1}N\:(\mathrm{mod}\,a)$. Note that $\overline{\overline{\imath}}=i$, and 
that $l^{(i)}$ exists if and only if $l^{(\overline{\imath})}$ exists with 
$l^{(i)}=l^{(\overline{\imath})}$.

We define \textit{degree} $\deg(F)\in\Phi(a)$ of $F\in K[\mathcal{X}]$ uniquely 
by $o(\deg(F))=o(F)$, and let $s:=\deg(F)$. From now on, $\Phi(a,o(s))$ is 
abbreviated to $\Phi(a,s)$.
Defining, for $l\in\Phi(a)$,
\begin{equation}\label{discrepancy}
dF_{l}:=\left\{\begin{array}{cl}
\sum_{n\in\Phi(a,s)}F_{n}u_{n+l^{(s_{2})}-s}
& \mathrm{if}\;l^{(s_{2})}\ge s,\\
0 & \mathrm{otherwise},
        \end{array}\right.
\end{equation}
where ``otherwise'' includes the vacant case of $l^{(s_{2})}$,
we call $dF_{l}$ \textit{discrepancy} of $F\in K[\mathcal{X}]$ at $l$. Let 
$V(u,N)$ be the set of $F\in K[\mathcal{X}]$ whose discrepancies are zero at 
all $l\in\Phi(a,N)$, and let $V(u,-1):=K[\mathcal{X}]$.
Then, for all $N\in\mathbb{Z}_{0}\cup\{-1\}$, $V(u,N)$ is an ideal in the 
ring $K[\mathcal{X}]$
(as proved at Proposition \ref{ideal} in Appendix \ref{V(u,A)}).
The BMS algorithm computes a Gr\"obner basis of $V(u,N)$ for each $N$, namely, a minimal polynomial ideal-basis with respect to the pole order $o(\cdot)$.
We may express the basis of $V(u,N)$ for each $N$ as $a$ polynomials 
$\{F_{N+1}^{(i)}(z)\}_{0\le i<a}$ by \eqref{expression}.
For sufficiently large $B$, we have $V(u,B)=I(\mathcal{E})$
(proved at Proposition \ref{ideal minimal set} in Appendix \ref{accord}).
Then $\{F_{B+1}^{(i)}(z)\}$ are called \textit{error-locator polynomials}, and 
the set of their common zeros agrees with $\mathcal{E}$.
Since the Goppa designed distance $d_{\mathrm{G}}$ of $\mathcal{C}(m)$ 
equals $m-2g+2$, we may set
\begin{equation}\label{m}
m:=2t+2g-1\quad
\mbox{for the correction up to $t$ errors},
\end{equation}
and can obtain $V(u,m)$ by using $\{u_{l}\}_{l\in\Phi(a,m)}$.

\begin{table}[t]
\centering
\caption{Values of $l^{(i)}=(l_{1}^{(i)},l_{2}^{(i)})\in\Phi^{(i)}(3,15)$
with $o(l^{(i)})=N$\label{l^(i)}}
  \resizebox{8.8cm}{!}{\includegraphics{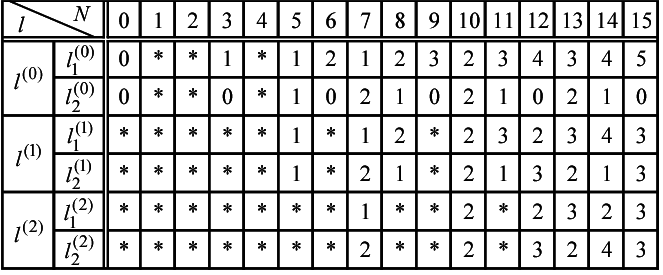}}
\end{table}

In the following inverse-free BMS algorithm, we denote the preserved 
condition (P) for updating formulae as follows: (P) $\Leftrightarrow$ 
$d_{N}^{(i)}=0$ or $s_{N}^{(i)}\ge l^{(i)}-c_{N}^{(\overline{\imath})}$.
\vspace{5mm}

\begin{description}
\item[{\bf Inverse-free BMS Algorithm}]
\item[Input]\quad syndrome values $\{u_{l}\}$ for $l\in\Phi(2a-1,m)$.
\item[Output]\quad error-locator polynomials $\{F^{(i)}_{m+1}(z)\}$.

\item[] In each step, the indicated procedures are carried out for all $0\le 
i<a$.
\newpage
\item[Step 0] (initializing)
$N:=0$,
$s_{N}^{(i)}:=(0,i)$,\\
$c_{N}^{(i)}:=(-1,i)$,
$v_{N}^{(i)}(Z):=\sum_{n\in\Phi(a,m)}u_{n}Z^{o(n)}$,\\
$w_{N}^{(i)}(Z):=1$, $f_{N}^{(i)}(Z):=1$, $g_{N}^{(i)}(Z):=0$.
\vspace{2mm}
\item[Step 1] (checking discrepancy)
If $l^{(i)}$ exists and $s_{N}^{(i)}\leq l^{(i)}$, then 
$d_{N}^{(i)}:=v_{N,N}^{(i)}$, else $d_{N}^{(i)}:=0$;\\
moreover, $e_{N}^{(i)}:=w_{N,N}^{(i)}$.
\vspace{2mm}
\item[Step 2] ($N$-updating)
\begin{align}
&s_{N+1}^{(i)}:=
\left\{\begin{array}{cl}
s_{N}^{(i)}&
\mathrm{if}\;\mathrm{(P)},\\
l^{(i)}-c_{N}^{(\overline{\imath})}&
\mathrm{otherwise},
        \end{array}\right.\label{ups}\\
&c_{N+1}^{(\overline{\imath})}:=
\left\{\begin{array}{cl}
c_{N}^{(\overline{\imath})}&
\mathrm{if}\;\mathrm{(P)},\\
l^{(i)}-s_{N}^{(i)}&
\mathrm{otherwise},
        \end{array}\right.\label{upc}\\
&f_{N+1}^{(i)}:=e_{N}^{(\overline{\imath})}f_{N}^{(i)}
-d_{N}^{(i)}g_{N}^{(\overline{\imath})}\label{upf},\\
&g_{N+1}^{(\overline{\imath})}:=
\left\{\begin{array}{cl}
Zg_{N}^{(\overline{\imath})}&
\mathrm{if}\;\mathrm{(P)},\\
Zf_{N}^{(i)}
& \mathrm{otherwise},
        \end{array}\right.\label{upg}\\
&v_{N+1}^{(i)}:=e_{N}^{(\overline{\imath})}v_{N}^{(i)}
-d_{N}^{(i)}w_{N}^{(\overline{\imath})}
\quad\mathrm{mod}\,Z^{N},\label{upv}\\
&w_{N+1}^{(\overline{\imath})}:=
\left\{\begin{array}{cl}
Zw_{N}^{(\overline{\imath})}&
\mathrm{if}\;\mathrm{(P)},\\
Zv_{N}^{(i)}
& \mathrm{otherwise}.
        \end{array}\right.\label{upw}
\end{align}

\item[Step 3] (checking termination)
If $N<m$, then $N:=N+1$ and go to Step 1, else stop the algorithm.
\hfill$\Box$
\end{description}
\vspace{6mm}

In the formula \eqref{upv},
``$\mathrm{mod}\,Z^{N}$'' means that
$v_{N+1}^{(i)}$ is defined by omitting
the term of $Z^{N}$ in $v_{N}^{(i)}$.
Then $v_{N}^{(i)}$, $w_{N}^{(i)}$ can be represented by
$$v_{N}^{(i)}(Z)=\sum_{h=N}^{m+N}v_{N,h}^{(i)}Z^{h},\quad
w_{N}^{(i)}(Z)=\sum_{h=N}^{m+N}w_{N,h}^{(i)}Z^{h},$$
and $v_{N,N}^{(i)}$, $w_{N,N}^{(i)}$ are defined by these.
We obtain $\{F^{(i)}_{N}(z)\}$ through
$$F_{N}^{(i)}(z):=\sum_{n\in\Phi(a,s)}f_{N,o(s-n)}^{(i)}z^{n}\quad\mathrm{with}\quad 
s:=s_{N}^{(i)}.$$
Then $d_{N}^{(i)}$ in the algorithm agrees with the discrepancy of 
$F_{N}^{(i)}$ at $o(l)=N$, i.e., $d_{N}^{(i)}=d(F_{N}^{(i)})_{l}$.

This inverse-free BMS algorithm is a novel version that eliminates the 
inverse calculation $\big(d_{N}^{(i)}\big)^{-1}$ from the parallel BMS 
algorithm \cite{IEEE05}\cite{AAECC-13}.
Compared with updating formulae in the original algorithm, which are 
later quoted at \eqref{oupf}--\eqref{oupw}, we see that 
\eqref{upf}--\eqref{upw} have eliminated the use of divisions, and in consequence have used $e_{N}^{(\overline{\imath})}$.
It is possible that one could remove the inverse calculation from the original (not parallel) BMS algorithm \textit{if} the values of 
$e_{N}^{(\overline{\imath})}$, which are actually previous values of 
$d_{N}^{(i)}$, are registered to memory-elements; in our parallel inverse-free BMS 
algorithm, we can conveniently take $e_{N}^{(\overline{\imath})}$ from the
coefficients of $w_{N}^{(\overline{\imath})}$ (as done in Step 1).

The following theorem confirms that 
$\{F_{N}^{(i)}\}_{0\le i<a}$ is a Gr\"obner basis of $V(u,N-1)$.

\begin{Theorem}\label{induction}
We have $F_{N}^{(i)}\in V(u,N-1)$,
$\mathrm{deg}(F_{N}^{(i)})=s_{N}^{(i)}$,
\begin{gather}\label{inequality}
s_{N,1}^{(0)}\ge s_{N,1}^{(1)}\ge\cdots\ge s_{N,1}^{(a-1)},
\;\mathrm{and}\\
\label{minimality}
s_{N,1}^{(i)}=\min\left\{\zeta_{N,1}^{(i)}\in{\mathbb Z}_{0}\left|
\begin{array}{l}
F\in V(u,N-1),\\
\mathrm{deg}(F)
=\left(\zeta_{N,1}^{(i)},i\right)
\end{array}\right.\right\}.\;\Box
\end{gather}
\end{Theorem}
The proof of Theorem \ref{induction} is referred to
Appendix \ref{Proof of Theorem},
in which $s_{N,1}^{(i)}=c_{N,1}^{(i)}+1$ is also obtained for all $N$ and 
$i$.

As explained at Proposition \ref{ideal minimal set} in Appendix 
\ref{accord}, the integer $B$ is required as $B\ge2t+4g-2+a$ to correct up 
to $t$ errors. Moreover, it is well-known 
\cite{Feng-Rao}\cite{Sakata-Jensen-Hoholdt} that the determination of 
unknown-syndrome values has to be done to proceed the loops for 
$N=m+1,m+2,\cdots,B$ of BMS algorithm.
In our Theorem \ref{induction}, as a result of division-less,
``$F_{N,s}^{(i)}=1$'' is not generally true 
differently from Theorem 1 of \cite{IEEE05},
and this fact disables us from generating the candidate values of unknown syndromes for majority voting.
Therefore, in our inverse-free BMS algorithm, we avoid the determination of unknown syndrome, and the loops of the algorithm are proceeded \textit{only} for $0\le N\le m$ by using the known syndrome values obtained directly from the 
received word.
Furthermore, we mainly consider the error-correction of generic errors 
\cite{independent}\cite{generic} (defined in the next section).
These techniques cause a slight decrease in the error-correcting 
capability; however, as described later in section \ref{Decoding of 
generic errors}, it does not matter in practice.

\begin{figure*}[!t]
\centering
  \resizebox{15.5cm}{!}{\includegraphics{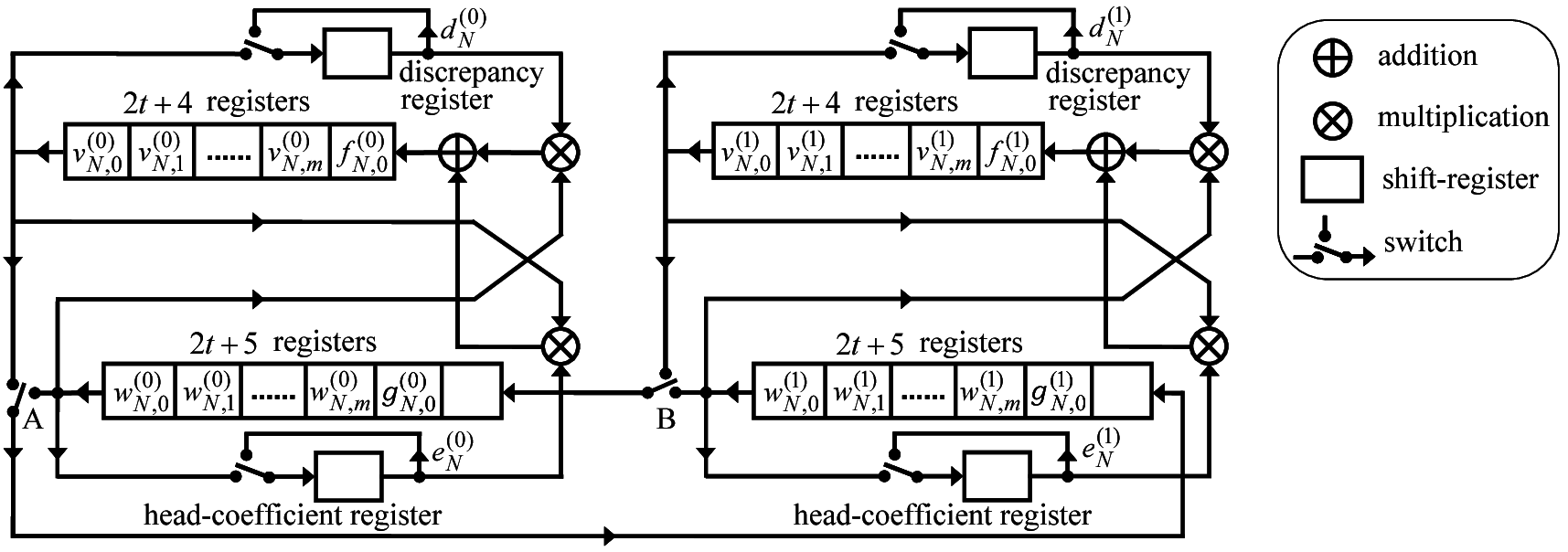}}
\caption{Inverse-free architecture for elliptic codes, which is composed of 
$a=2$ blocks exchanging $w_{N}^{(i)}$ and $g_{N}^{(i)}$.\label{inverse-free 
decoder}}
\end{figure*}
\begin{figure*}[!t]
\centering
  \resizebox{16cm}{!}{\includegraphics{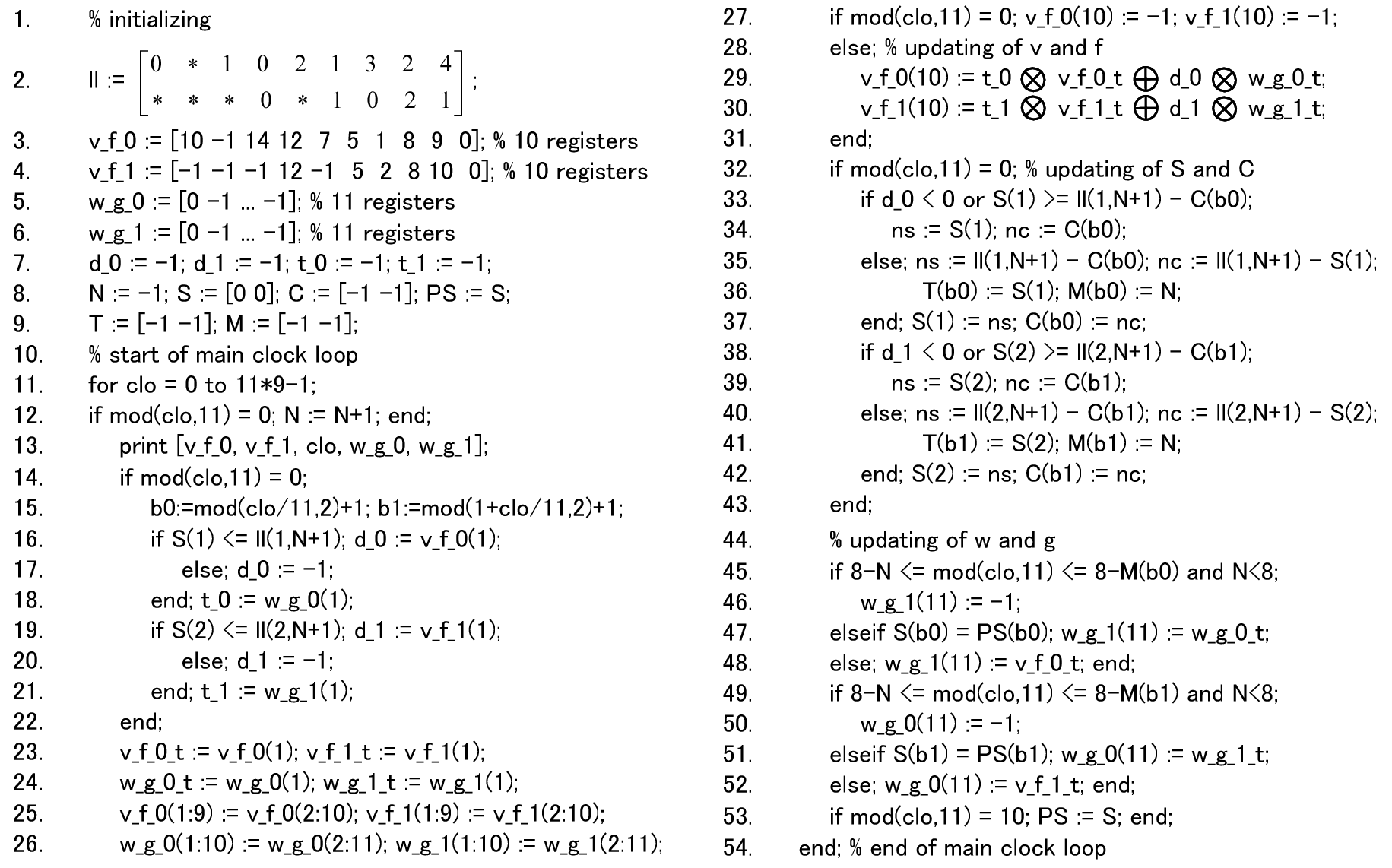}}
\caption{Program simulating the inverse-free architecture for $(24,16,8)$ elliptic code $\mathcal{C}(8)$ over GF($2^4$) with three-error correction.\label{program}}
\end{figure*}

\section{Inverse-free architecture}\label{Inverse-free architecture}
As the first of three kinds of architectures proposed in this paper, we 
describe \textit{inverse-free architecture}, which has the plainest structure 
of the three.
\vspace{-2mm}

\subsection{Model structure}\label{Model structure}
In this subsection, we give a direct application of the inverse-free BMS algorithm,
which corresponds to K\"otter's architecture \cite{K"otter} of which inverse-calculators have been replaced by multipliers.
To make the case clear, we describe the architecture for elliptic codes, 
that is, codes on elliptic curves, although we take the generality into 
account; we can employ it for other codes on algebraic curves without 
difficulty.

As shown in the model Fig. \ref{inverse-free decoder}, the coefficients of 
$v_{N}^{(i)}$, $f_{N}^{(i)}$ are arranged in a sequence of shift-registers, 
and those of $w_{N}^{(i)}$, $g_{N}^{(i)}$ are arranged in another sequence.
It is similar to K\"otter's architecture \cite{K"otter} that the proposed 
architecture has $a$-multiple structure (i.e. $a$ blocks) of the 
architecture for the Berlekamp--Massey algorithm \cite{Berlekamp}\cite{Massey} of RS codes.
The difference is that $a$ division-calculators in the K\"otter's 
architecture are replaced with $a$ multipliers in our architecture.
Moreover, while the values of discrepancy are computed in the K\"otter's architecture with one multiplier and a shift-register according to definition 
\eqref{discrepancy}, our architecture derives the values from the 
coefficients of $v_{N}^{(i)}$ with \textit{discrepancy registers} and reduces
the one multiplier for computing discrepancy.

In Fig. \ref{inverse-free decoder}, we omit input and output terminals, 
and the initial ($N=0$) arrangement of the coefficients in polynomials is 
indicated. The number of registers in one shift-register sequence for 
$v_{N}^{(i)}$ and $f_{N}^{(i)}$ should be equal to the total number of 
coefficients in $v_{N}^{(i)}$ and $f_{N}^{(i)}$, i.e., $m+2$ for 
$\mathcal{C}(m)$;
although it might seem that there is no space for $f_{N}^{(i)}$,
it is made by shortening and shifting
of $v_{N}^{(i)}$ as $N$ is increased.
On the other hand, the number of shift-registers required for $w_{N}^{(i)}$ and $g_{N}^{(i)}$ is one more than that for 
$v_{N}^{(i)}$ and $f_{N}^{(i)}$ because of the structure of parallel BMS 
algorithm, and should be $m+3$.

If $N\equiv0\:\mathrm{mod}\,(m+3)$, the switches in the discrepancy 
registers are closed downward to obtain the values of discrepancy 
$v_{N,N}^{(i)}=d_{N}^{(i)}$, and if $N\not\equiv0\:\mathrm{mod}\,(m+3)$, 
they are closed upward to output the values of discrepancy at each clock.
The head-coefficient registers work similarly to the discrepancy registers, 
and output the values of the head coefficient $w_{N,N}^{(i)}=e_{N}^{(i)}$ of 
$w_{N}^{(i)}$.
The coefficients of $w_{N}^{(i)}$ and $g_{N}^{(i)}$ are transferred from 
the block of $v_{N}^{(\overline{\imath})}$ to that of $v_{N+1}^{(\overline{\imath})}$ $(\overline{\imath}$ for $N+1)$.
The switches A and B work according to the preserving or updating of 
$w_{N}^{(i)}$ and $g_{N}^{(i)}$, i.e., ``(P)'' or ``otherwise'' in 
\eqref{upg} and \eqref{upw}.

\begin{table*}[!t]
\caption{Values of registers in four shift-register sequences, discrepancy 
$d_{N}^{(i)}$, and $s_{N,1}^{(i)}$ in the inverse-free 
architecture.\label{register value if}}
\centering
  \resizebox{16.8cm}{!}{\includegraphics{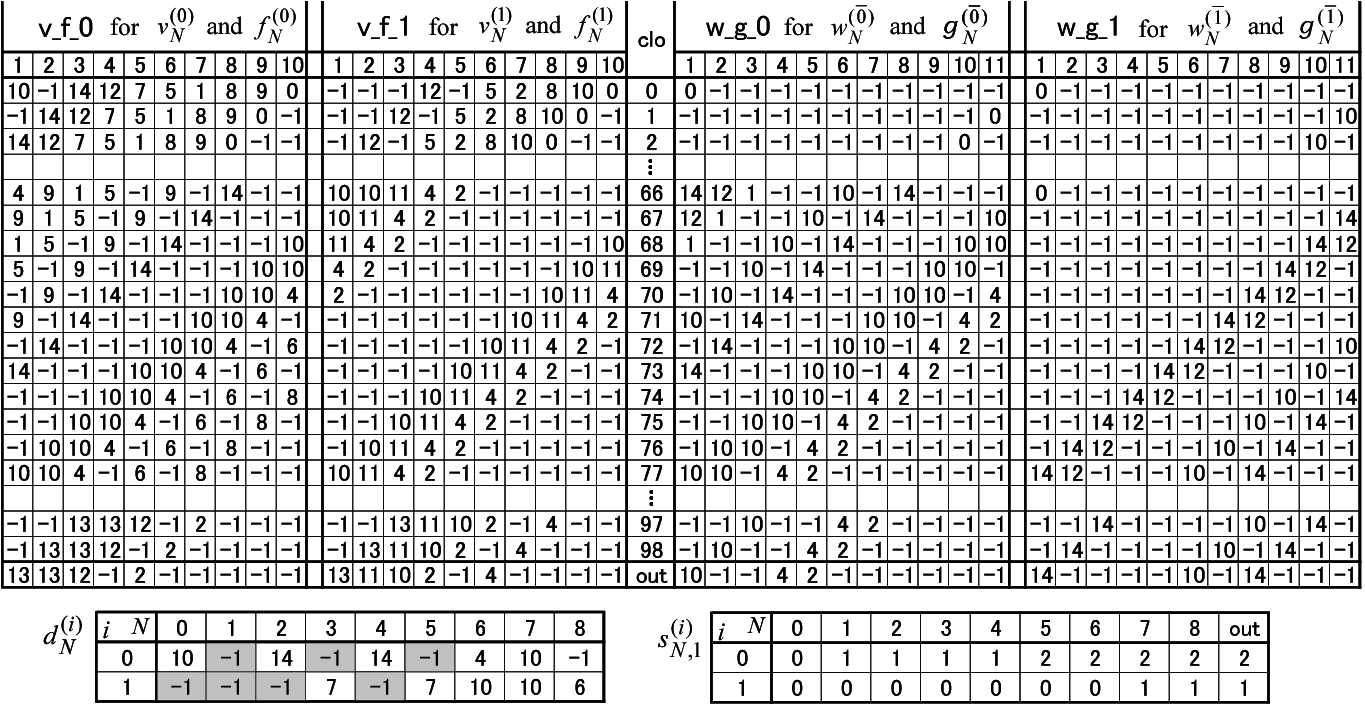}}
\end{table*}

Thus, one may only perform simple additions and multiplications for the 
values in the shift-register sequences for $v_{N}^{(i)}$ and $f_{N}^{(i)}$ 
to update them.
On the other hand, as for $w_{N}^{(i)}$ and $g_{N}^{(i)}$, one must not only 
perform additions and multiplications but also set register-values to zero, 
or else old disused values corrupt $v_{N}^{(i)}$ and $f_{N}^{(i)}$.
We describe this procedure in a later subsection \ref{Simulation}.

This inverse-free architecture has an $a$-multiple structure closer to 
K\"otter's than to the latter two architectures, and has been changed to 
division-free and parallel in the sense of using two types of polynomials,
$v_{N}^{(i)}$ and $w_{N}^{(i)}$, to compute discrepancy.
We see in Section \ref{Performance estimation} that the total number of 
shift-registers in our architecture is nearly the same as that in K\"otter's, i.e., the additional polynomials do not contribute essentially to the total 
number of registers.

\subsection{Decoding of generic errors}\label{Decoding of generic errors}
To implement the inverse-free algorithm effectively, we concentrate on 
decoding generic $t$-errors \cite{independent}\cite{generic}, for which the 
degree $s_{N}^{(i)}$ of error-locator polynomials is characterized by 
$o(s_{N}^{(i)})\le t+g-1+a$, while in general we have $o(s_{N}^{(i)})\le 
t+2g-1+a$.
In other word, the error-location $\mathcal{E}$ is generic if and only if so-called \textit{delta set} $\{l\in\Phi(a)\,|\,l\le s_{N}^{(l_{2})}\}$ of error-locator polynomials corresponds to the first $t$ non-gaps in $o\left(\Phi(s)\right)$.
Then the loops of BMS algorithm are required for $0\le N\le m+a-1$ to obtain 
the error-locator polynomials for generic $t$-errors, while in general $0\le 
N\le m+2g-1+a$ for all errors;
these facts are proved in Appendix \ref{Generic case}.
Thus we see that $\left(t-\left\lceil(a-1)/2\right\rceil\right)$ errors are 
corrected in $\mathcal{C}(m)$ after $N$-updating for $0\le N\le m$.
The merits of this method are not only that it is inverse-free and there is no majority logic \cite{Feng-Rao} but also that there are fewer loops of the BMS algorithm; we can cut it down to $2g-1$ loops. Furthermore, this method can also be applied to K\"otter's and systolic-array architectures \cite{IEEE05}.

There are two drawbacks to this method.
The first is that non-generic errors cannot be corrected.
Since generic or non-generic is also defined by whether a matrix determinant $\not=0$ or not (as shown in Appendix \ref{Generic case}), the ratio of generic errors to all 
errors is estimated at $(q-1)/q$, under the hypothesis for the randomness of 
values $\{z^{l}(P_{j})\}$ (which is supported by numerical tests 
\cite{ISITA04}).
As for a practical size $q=2^8$, the ratio is equal to 
$255/256=0.9960\cdots$. Moreover, for errors less than $t$, the percentage of correctable errors increases since $o(s_{N}^{(i)})$s decrease.
Thus we have less effect of this drawback.
The second is that the number of correctable errors is decreased 
$\left\lceil(a-1)/2\right\rceil$ for $t$-error correctable codes 
$\mathcal{C}(m)$.
This corresponds to $t-1$ errors for all elliptic codes, and $t-8$ errors 
for Hermitian codes over $\mathbb{F}_{2^8}$.
However, this has no serious effect on practical function;
we might choose $\mathcal{C}(m+a-1)$ to correct $t$ errors, and the 
remaining error-correcting capability is available for error-detection up to 
$t+\left\lfloor(a-1)/2\right\rfloor$ errors.
In the next subsection, we demonstrate the decoding of $\mathcal{C}(m)$ with 
$m:=m+1$ (i.e. $a=2$) for $t$-error correction in codes on elliptic curves.

\begin{figure*}[!t]
\centering
  \resizebox{13cm}{!}{\includegraphics{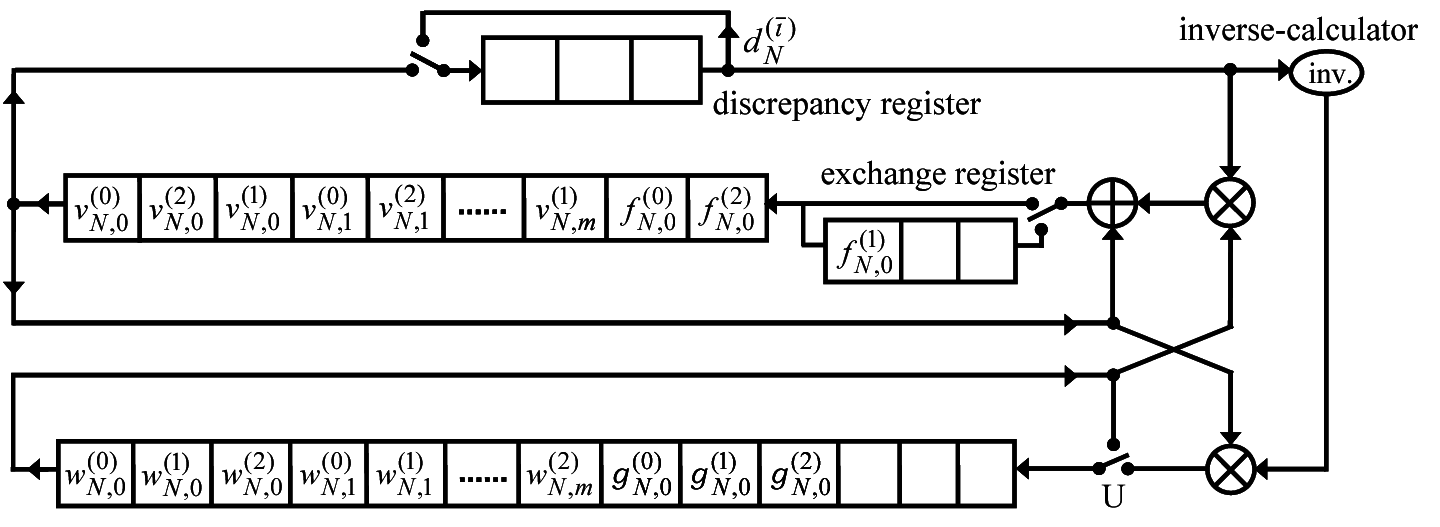}}
\caption{Serial architecture for Klein-quartic codes, which has a single 
structure with serially-arranged coefficients.\label{serial decoder}}
\end{figure*}

\subsection{Simulation and numerical example}\label{Simulation}
In this subsection, we focus on an elliptic code, especially on the elliptic 
curve defined by the equation $y^{2}+y=x^{3}+x$ over $K:={\mathbb F}_{16}$, 
and simulate a decoder for it. This curve has 25 $K$-rational points equal 
to the Hasse-Weil bound with genus one, and we obtain code $\mathcal{C}(m)$ 
of length 24.

We choose a primitive element $\alpha$ of $K$ satisfying 
$\alpha^{4}+\alpha=1$, and represent each non-zero element of $K$ as the 
number of powers of $\alpha$. Moreover, we represent zero in $K$ as $-1$; 
note that, e.g., 0 and $-1$ mean $1=\alpha^{0}$ and 0, respectively. Let the 
set of error-locations $\mathcal{E}:=\{(x,y)=(3,7),(9,11),(14,4)\}$, and let the error-values be 6, 8, 11, respectively.

In Fig. \ref{program}, we provide a brief description of MATLAB m-file 
program for our architecture, where $\mathrm{mod}(x,Y)$ returns the smallest 
non-negative integer satisfying $x\equiv\mathrm{mod}(x,Y)$ 
$(\mathrm{mod}\,Y)$.
Comments are written next to ``{\sf\%}.''
At line {\sf2}, {\sf ll}$(1+i,1+N)$, which corresponds to the $(1+i,1+N)$-th 
component of matrix {\sf ll} in MATLAB m-file notations, defines 
$l_{1}^{(i)}$ with $N=o(l^{(i)})$ of $l^{(i)}\in\Phi^{(i)}(2,8)$ to decode 
$3$ errors in $\mathcal{C}(8)$ with $m=8$. In the case $l_{1}^{(i)}=*$ in 
{\sf ll}, the logical sentences at lines {\sf 16} and {\sf 19} are regarded 
to be false.

In the case of elliptic codes $\mathcal{C}(m+1)$,
the number of registers for $v_{N}^{(i)}$ and $f_{N}^{(i)}$ should be 
$(m+1)+2=2t+4$ by \eqref{m}, and that for $w_{N}^{(i)}$ and $g_{N}^{(i)}$ 
should be $2t+5$, as in lines {\sf3}--{\sf6} for $t=3$.
At line {\sf 15}, the value {\sf b0} (resp. {\sf b1}) corresponds to 
$\overline{\imath}$ at $N$ for $i=0$ (resp. $i=1$).
At lines {\sf 25} and {\sf 26}, the shift-register values are shifted to the 
neighbors, and, e.g., ``{\sf v\underline{ }f\underline{ }0(1:9):=v\underline{ }f\underline{ }0(2:10)}'' indicates the shifts of nine values {\sf v\underline{ }f\underline{ }0(1):=v\underline{ }f\underline{ }0(2)}, $\cdots$, {\sf v\underline{ }f\underline{ }0(9):=v\underline{ }f\underline{ }0(10)}, where {\sf v\underline{ }f\underline{ }0}$(n)$ corresponds to the $n$-th component 
of {\sf v\underline{ }f\underline{ }0}.

Table \ref{register value if} shows that our architecture outputs the 
error-locator polynomials $\{F_{m+1}^{(i)}(z)\}$ and the auxiliary 
polynomials $\{G_{m+1}^{(i)}(z)\}$ for $\mathcal{E}$. The top of Table 
\ref{register value if} indicates the indexes of registers of four 
shift-register sequences. The center column indicates the values of ``{\sf 
clo}'' in the program, which corresponds to the underlying clock of the 
architecture.
The values of discrepancy $d_{N}^{(i)}$ are indicated at the left bottom of 
Table \ref{register value if}, where
``$\,\resizebox{8mm}{!}{\includegraphics{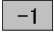}}\,$'' indicates 
the state that $l^{(i)}$ does not exist or $s_{N,1}^{(i)}>l_{1}^{(i)}$.
The values of discrepancy $d_{N}^{(i)}$ are obtained at {\sf clo} $=11N$
from {\sf v\underline{ }f\underline{ }0(1)} or {\sf v\underline{ 
}f\underline{ }1(1)} if $s_{N,1}^{(i)}\le l_{1}^{(i)}$.
The values of $s_{N}^{(i)}$ are indicated at the right bottom of Table 
\ref{register value if}.

\begin{figure*}[!t]
\centering
  \resizebox{16cm}{!}{\includegraphics{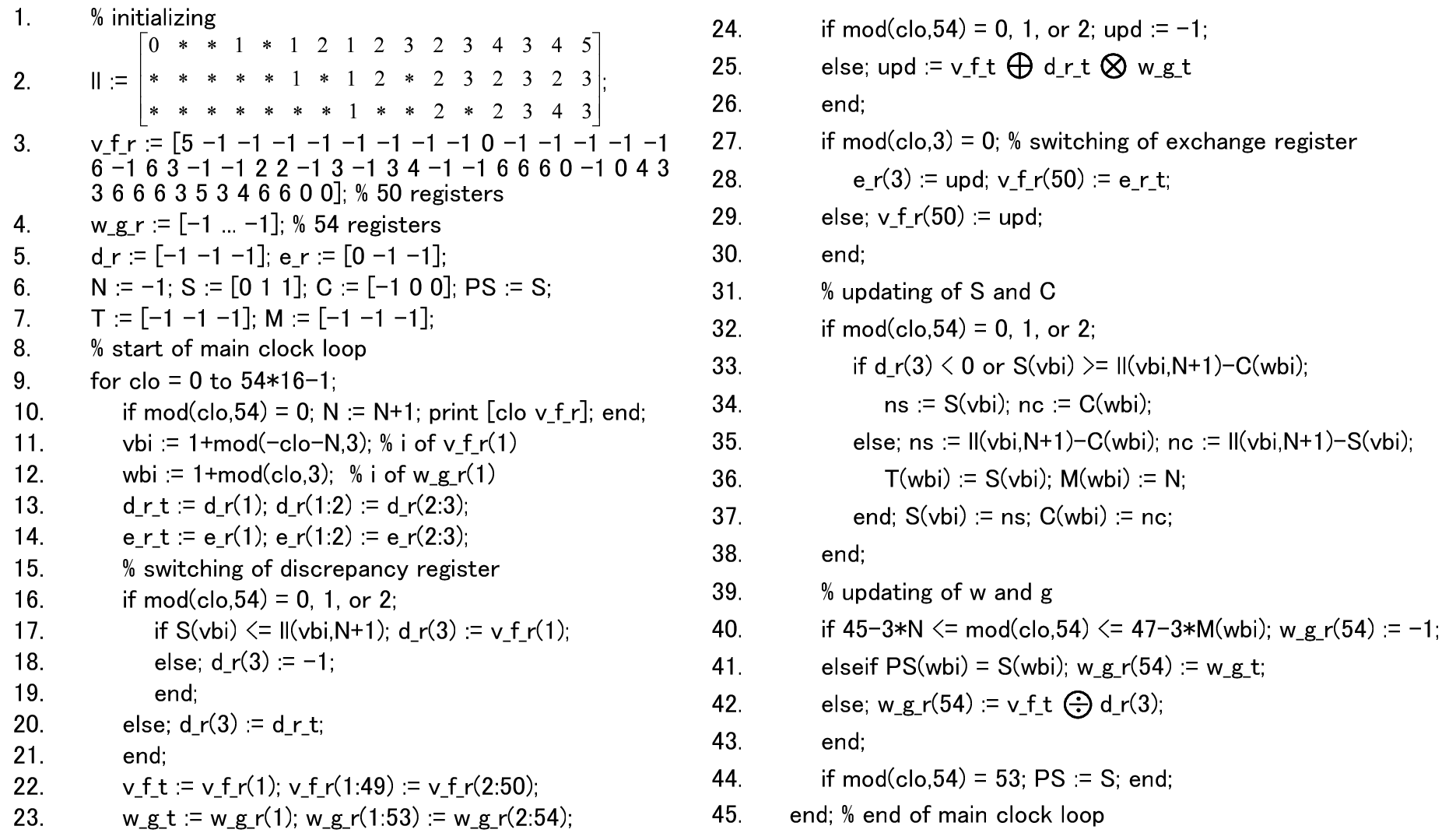}}
\caption{Program simulating the serial architecture for $(23,10,11)$ code $\mathcal{C}(15)$ on Klein's quartic over GF($2^3$) with four-error correction.\label{serial program}}
\end{figure*}

The most difficult point in the program is that suitable register values 
must be settled to $-1$ at the lines {\sf45} and {\sf49} for not changing the coefficients of $f_{N}^{(i)}$. Let $t_{N}^{(i)}:=\deg(G_{N}^{(i)}(z))$ and 
$M^{(i)}$ be the value of $N$ at which the last updating of $G_{N}^{(i)}$ 
occurred; we have $t_{N}^{(i)}=s_{M^{(i)}}^{(\overline{\imath})}$ with 
$\overline{\imath}$ at $M^{(i)}$, and have $t_{N,1}^{(i)}=$ {\sf T}$(1+i)$, 
$M^{(i)}=$ {\sf M}$(1+i)$ in the program.
\textit{Then, we claim that $g_{N,N-M^{(i)}}^{(i)}$, that is, the head 
coefficient of
$$
g_{N}^{(i)}=\sum_{h=N-M^{(i)}}^{o(t_{N}^{(i)})+N-M^{(i)}}g_{N,h}^{(i)}Z^{h}
$$
is located at the $(10-M^{(i)})$-th register of {\sf w\underline{ 
}g\underline{ }0} or {\sf w\underline{ }g\underline{ }1} according to 
$\overline{\imath}=0$ or $1$ if {\sf mod(clo,11)} $=0$.}
For example, if {\sf clo} $=66$ and {\sf N} $=6$, we can see from 
$s_{N,1}^{(i)}$ in Table \ref{register value if} that $M^{(0)}=4$. Then 
$g_{6,2}^{(0)}=\alpha^{10}$ is in {\sf w\underline{ }g\underline{ }0(6)}. As 
another example, if {\sf clo} $=77$ and {\sf N} $=7$, we can see that 
$M^{(1)}=6$, and then $g_{7,1}^{(1)}=\alpha^{4}$ is in {\sf w\underline{ 
}g\underline{ }0(4)}.

Noting that the value in {\sf w\underline{ }g\underline{ }0(j)} at {\sf 
mod(clo,11)} $=0$ is the shifted value at {\sf mod(clo,11)} $=j-1$, e.g.,
{\sf w\underline{ }g\underline{ }0(11) := w\underline{ }g\underline{ }1(1)}, 
we obtain the upper and lower conditions of {\sf w\underline{ }g\underline{ 
}0(11)} and {\sf w\underline{ }g\underline{ }1(11)} $:=-1$ at lines {\sf45} 
and {\sf49}, since each $N+1-M^{(i)}$ value of {\sf w\underline{ 
}g\underline{ }0(j)} and {\sf w\underline{ }g\underline{ }1(j)} for $j=9-N$, 
$9-N+2$, $\cdots$, $9-M^{(i)}$ must be $-1$ at {\sf mod(clo,11)} $=0$ in 
each $w_{N}^{(i)}$.
The condition ``{\sf N$<$8}'' is required to obtain the values of 
$e_{9}^{(i)}:=w_{9,9}^{(i)}$ for error-evaluation (stated below).

Thus, the Gr\"obner basis 
$\{F_{9}^{(0)}=\alpha^{13}x^{2}+\alpha^{13}y+\alpha^{12}x+\alpha^{2},\,F_{9}^{(1)}=\alpha^{13}xy+\alpha^{11}x^{2}+\alpha^{10}y+\alpha^{2}x+\alpha^{4}\}$ 
of ideal $I(\mathcal{E})$ has been obtained together with the auxiliary 
polynomials 
$\{G_{9}^{(0)}=\alpha^{10}x+\alpha^{14},\,G_{9}^{(1)}=\alpha^{4}y+\alpha^{2}x\}$.
We obtain the set $\mathcal E$ of error-locations through the Chien search,
and obtain each error-value by O'Sullivan's formula
\cite{O'Sullivan error values}
\begin{equation}\label{O'Sullivan's formula}
e_{j}=\Bigg(\sum_{0\le i<a}
\frac{F_{m+1}^{(i)\,\prime}(P_{j})}{F_{m+1,s}^{(i)}}
\frac{G_{m+1}^{(i)}(P_{j})}{e_{m+1}^{(i)}}
\Bigg)^{-1}\;\mathrm{for}\;P_{j}\in\mathcal{E},
\end{equation}
where $F_{m+1}^{(i)\,\prime}(z)$ is
the formal derivative of $F_{m+1}^{(i)}(z)$
with respect to $x$, e.g., $y^{\,\prime}=x^{2}+1$.
Note that the divisions in this formula are independent from BMS algorithm,
and are calculated by the repetitional multiplications using the multipliers 
in our architecture as follows.

Since we have $\beta^{-1}=\beta^{2^{n}-2}$ for 
$0\not=\beta\in\mathbb{F}_{2^{n}}$, and have $a_{n}=2^{n}-1$ for the 
sequence defined by $a_{1}:=1$ and $a_{n+1}:=2a_{n}+1$, we see that the 
calculation of $\beta^{-1}$ consists of $(n-2)$ multiplications of $\beta$ 
and $(n-1)$ squares, and the total is $(2n-3)$ multiplications in 
$\mathbb{F}_{2^{n}}$. Thus we can say that our architecture eliminates $a$ 
inverse-calculators, each of which corresponds to $(2n-3)$ multipliers, with 
$\left\lfloor\frac{a-1}{2}\right\rfloor$ slight drop of error-correction 
capability for $\mathcal{C}(m+a-1)$.

\section{Serial architecture}\label{Serial architecture}
As the second architecture, we describe \textit{serial architecture} \cite{SITA04}, which has 
a different structure from K\"otter's and the preceding ones.
In this section, we focus on well-known codes on Klein's quartic curve over 
$K:={\mathbb F}_{8}$, and simulate a decoder for it.
Many articles so far have treated codes on this curve as examples.

Klein's quartic curve is defined by equation $X^{3}Y+Y^{3}Z+Z^{3}X=0$ in 
projective plane $\mathbb{P}^2=\{(X:Y:Z)\}$, which causes
$y^{3}x+x^{3}+y=0$ by $(x,y):=(Y/Z,X/Z)$ in the affine form, and has the 
same number of $K$-rational points as Hasse--Weil--Serre upper bound 24 with 
genus 3.
We denote $K$-rational points $(X:Y:Z)=(1:0:0)$ and $(0:1:0)$ as 
$P_{(1:0:0)}$ and $P_{(0:1:0)}$, and other 22 points as the values of 
$(x,y)$. Although it is not a C${}_{a}^{b}$ curve, the monomial basis of 
$L(mP_{(0:1:0)})$ to make $\mathcal{C}(m)$ is obtained by
$\{x^{n_{1}}y^{n_{2}}\,|\,n\in\Phi(3,m)\}\backslash\{y,y^2\}$
with $o(n):=3n_{1}+2n_{2}$ and the minimal pole order $a=3$ as in Fig. 
\ref{Phi(3,15)}. We note that $x(P_{(1:0:0)})=(xy)(P_{(1:0:0)})=0$ and 
$(xy^2)(P_{(1:0:0)})=1$, and then obtain code $\mathcal{C}(m)$ of length 23.

We intend to correct generic errors in $\mathcal{C}(m+2)$ with $m:=2t+5$ 
(cf. \ref{Decoding of generic errors}).
Let a primitive element $\alpha$ of $K$ be $\alpha^{3}+\alpha=1$. We 
represent each non-zero element of $K$ as the number of powers of $\alpha$ 
as in \ref{Simulation}. Let the set of error-locations 
$\mathcal{E}:=\{(x,y)=(0,1),(1,0),(2,0),(3,3)\}$, and let error-values be 1, 
2, 5, 4, respectively.

As in the model Fig. \ref{serial decoder}, the serial architecture has a 
single structure similar to that of RS codes, while K\"otter's and the 
preceding inverse-free architectures have an $a$-multiple structure.
The initial ($N=0$) arrangement of the coefficients in polynomials is also 
indicated in Fig. \ref{serial decoder}.
In the case of the architecture for codes on Klein's quartic, it is convenient to exchange $i$ and $\overline{\imath}$ in all updating formulae 
\eqref{ups}--\eqref{upw}, and the validity follows from 
$\overline{\overline{\imath}}=i$.
For the serial architecture, we employ not the inverse-free BMS algorithm but the original parallel BMS algorithm \cite{IEEE05}\cite{AAECC-13},
which is described by exchanging updating formulae 
\eqref{upf}--\eqref{upw} into the following (quoted from \cite{IEEE05}):
\vspace{-2mm}
\begin{align}\label{oupf}
&f_{N+1}^{(\overline{\imath})}:=f_{N}^{(\overline{\imath})}
-d_{N}^{(\overline{\imath})}g_{N}^{(i)},\\
&g_{N+1}^{(i)}:=\label{oupg}
\left\{\begin{array}{cl}
Zg_{N}^{(i)}&
\mathrm{if}\;\mathrm{(P)},\\
\big(d_{N}^{(\overline{\imath})}\big)^{-1}Zf_{N}^{(\overline{\imath})}
& \mathrm{otherwise},
        \end{array}\right.\\
&v_{N+1}^{(\overline{\imath})}:=v_{N}^{(\overline{\imath})}
-d_{N}^{(\overline{\imath})}w_{N}^{(i)}\label{oupv}
\quad\mathrm{mod}\,Z^{N},\\
&w_{N+1}^{(i)}:=\label{oupw}
\left\{\begin{array}{cl}
Zw_{N}^{(i)}&
\mathrm{if}\;\mathrm{(P)},\\
\big(d_{N}^{(\overline{\imath})}\big)^{-1}Zv_{N}^{(\overline{\imath})}
& \mathrm{otherwise}.
        \end{array}\right.
\end{align}
\vspace{-2mm}

Then the coefficients of $v_{N}^{(\overline{\imath})}$ and 
$f_{N}^{(\overline{\imath})}$ are arranged \textit{serially} in the order 
$\overline{\imath}=0,2,1$ in one sequence of shift-registers, and those of 
$w_{N}^{(i)}$ and $g_{N}^{(i)}$ are arranged in the order $i=0,1,2$
in another.
This arrangement of coefficients is decided by the pair 
$(\overline{\imath},\,i)$, and is special to the codes on Klein's quartic; for codes on C${}_{a}^{b}$ curves, see the next subsection.

Instead of the round of $\{w_{N}^{(i)},g_{N}^{(i)}\}$ $(0\le i<a)$ among $a$ 
blocks in the preceding architecture, the order $\overline{\imath}=0,2,1$ of 
$\{v_{N}^{(\overline{\imath})},f_{N}^{(\overline{\imath})}\}$ at 
$N\equiv0\:(\mathrm{mod}\,a)$ is changed to $\overline{\imath}=2,1,0$ at 
$N\equiv1$, and to $1,0,2$ at $N\equiv2$, and so on.
Although one may change the order of the coefficients of 
$\{w_{N}^{(i)},g_{N}^{(i)}\}$, our layout is easier because of the existence 
of updating (i.e., the switch ``U'' in Fig. \ref{serial decoder}).

\begin{table*}[!t]
\caption{Values of registers in two shift-register sequences, discrepancy 
$d_{N}^{(i)}$, and $s_{N,1}^{(i)}$ in the serial architecture.\label{serial 
register value}}
\centering
  \resizebox{17.3cm}{!}{\includegraphics{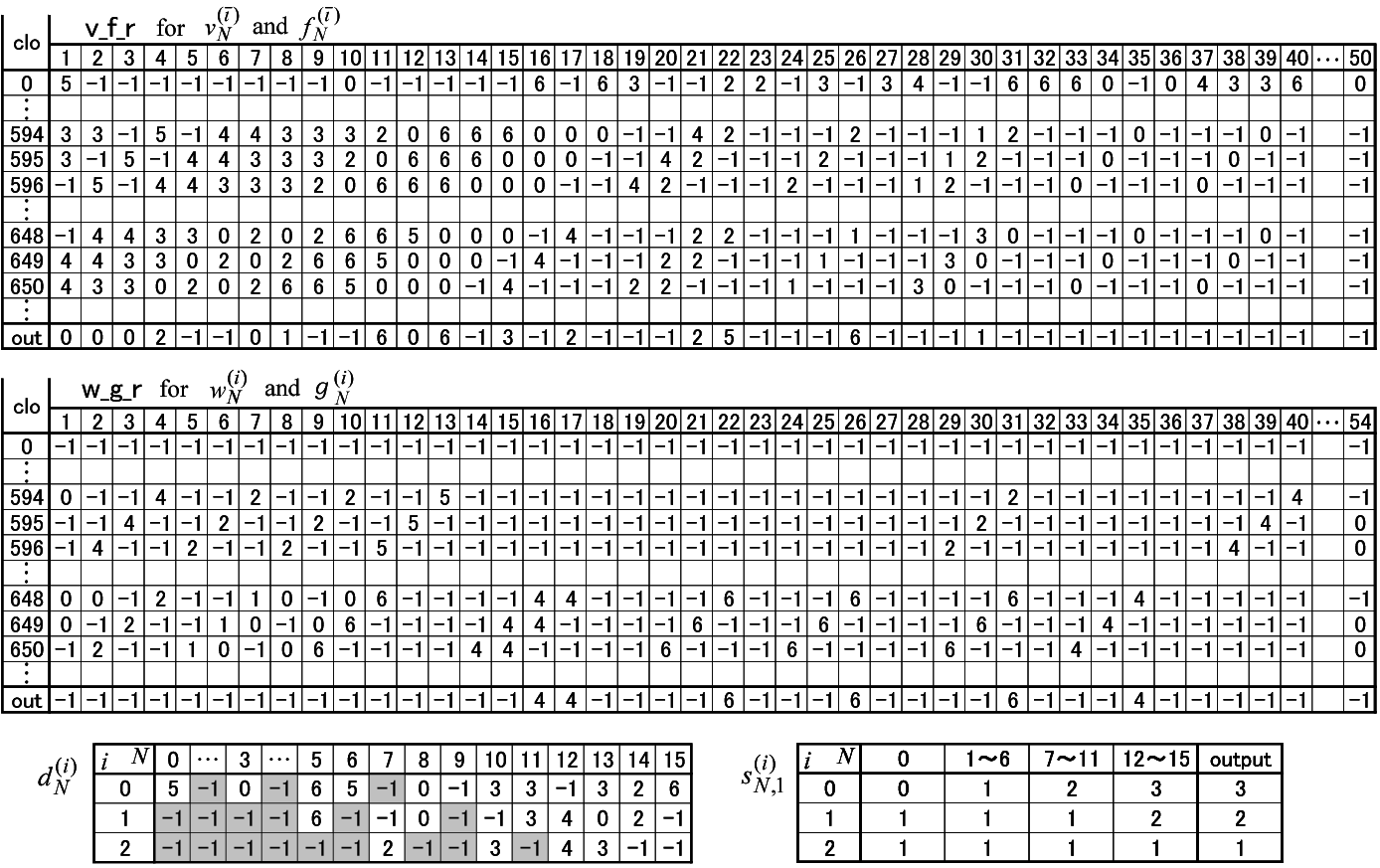}}
\end{table*}

The exchange register has this role of changing the order.
We introduce a method to carry it out with only shift-registers and 
switches. The following is a small example; at 
$\mathrm{mod}(\mathrm{clo},3)=0$, the switch is down to take the leftmost 
value in the exchange register, and at other clo's, the switch is up in 
order to pass it.
$$\resizebox{6.5cm}{!}{\includegraphics{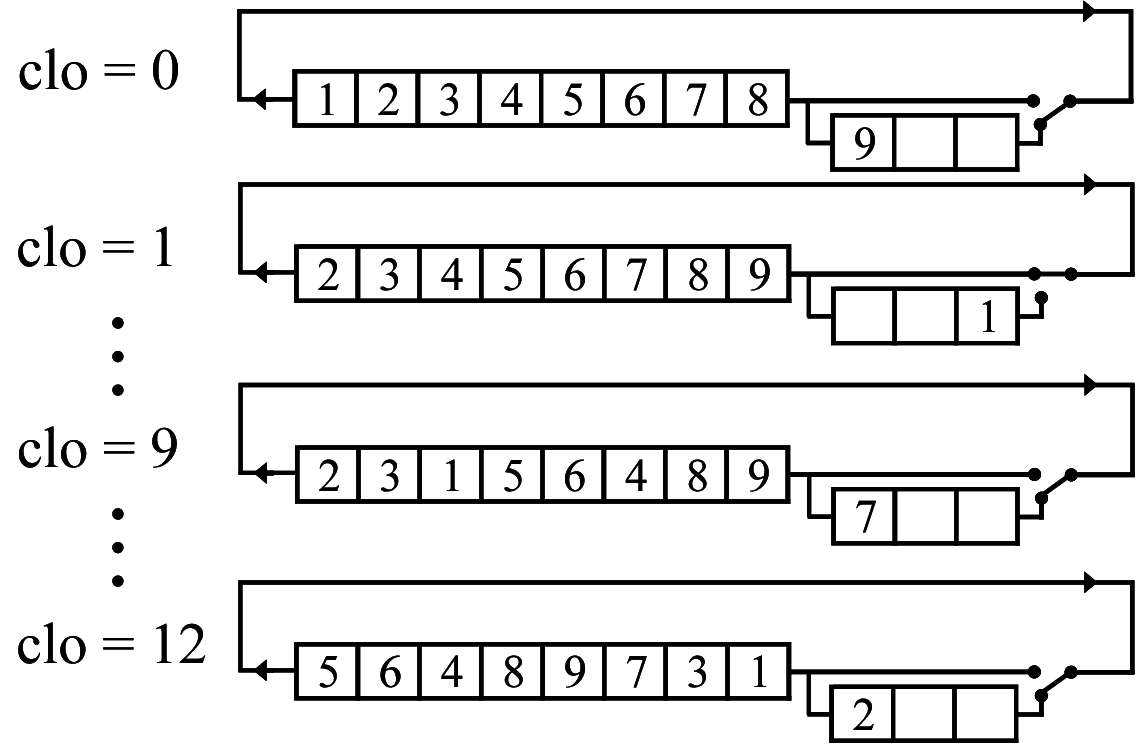}}$$
We can see that the exchange register works like a shift-register,
since the order-changing has been finished at $\mathrm{clo}=9$ and the 
omission by $\mathrm{mod}\,Z^{N}$ in \eqref{upv} has been done after $a$ 
more clo's.

The number of registers in one shift-register sequence for $v_{N}^{(i)}$s 
and $f_{N}^{(i)}$s should be equal to the total number of coefficients 
minus one, i.e., $3(m+2)-1$ for $\mathcal{C}(m)$, and this works like 
$3(m+2)$ together with the exchange registers.
On the other hand, $w_{N}^{(i)}$s and $g_{N}^{(i)}$s require $a$ more shift-registers than $v_{N}^{(i)}$s and $f_{N}^{(i)}$s
because of the structure of parallel BMS algorithm.
Thus the number of registers for $w_{N}^{(i)}$s and $g_{N}^{(i)}$s should 
be $3(m+2)+3$.
Then $6t+26$ and $6t+30$ registers are required for $\mathcal{C}(m+2)$ with 
$m=2t+5$.

In Fig. \ref{serial program},
we describe the architecture with a MATLAB m-file program, where the notations are the same as in Fig. \ref{program}.
At line {\sf6}, the values of $[s_{N,1}^{(0)},s_{N,1}^{(1)},s_{N,1}^{(2)}]$ 
and $[c_{N,1}^{(0)},c_{N,1}^{(1)},c_{N,1}^{(2)}]$ are initialized
differently from all 0 and $-1$ because of the exclusion of 
$\{(0,1),(0,2)\}$ from $\Phi(3)$.

The most difficult point in the program is again that suitable register 
values should be settled to zero at line {\sf 40} in the successive loop 
for not meeting the coefficients of $f_{N}^{(i)}$.
Since $\alpha^{0}=f_{0,0}^{(0)}$ is at the 49-th register in the initial 
values of {\sf v\underline{ }f\underline{ }r}, we claim that 
$g_{N,N-M^{(i)}}^{(i)}$ (the head coefficient of $g_{N}^{(i)}$) is located 
at the $(49-3M^{(i)})$-th register of {\sf w\underline{ }g\underline{ }r} if 
{\sf mod(clo,54)} $=i$. For example, if {\sf clo} $=648$ and $N=12$, we can 
see from $s_{N,1}^{(i)}$ in Table \ref{serial register value} that 
$M^{(0)}=M^{(1)}=11$. Then $g_{12,1}^{(0)}=g_{12,1}^{(1)}=\alpha^{4}$ are in 
{\sf w\underline{ }g\underline{ }r(16)} at {\sf clo} $=648$ and 649.

Similarly as in Subsection \ref{Simulation}, we note that the value in {\sf 
w\underline{ }g\underline{ }r(j)} at {\sf mod(clo,54)} $=i$ is the shifted 
value at {\sf mod(clo,54)} $=i+j-1$, e.g., {\sf w\underline{ }g\underline{ 
}r(54) := v\underline{ }f\underline{ }r(1)}.
Moreover, since each $N+1-M^{(i)}$ value of {\sf w\underline{ }g\underline{ 
}r(j)} for $j=46-3N$, $46-3N+3$, $\cdots$, $46-3M^{(i)}$ must be $-1$ at 
{\sf mod(clo,54)} $=i$ in each $w_{N}^{(i)}$,
we obtain the upper and lower conditions of {\sf w\underline{ }g\underline{ 
}r(54)} $:=-1$ at line {\sf40} as the union of
\begin{align*}
i&=0\;\Rightarrow\;j=45-3N,\,\cdots,\,45-3M^{(0)},\\
i&=1\;\Rightarrow\;j=46-3N,\,\cdots,\,46-3M^{(1)},\\
i&=2\;\Rightarrow\;j=47-3N,\,\cdots,\,47-3M^{(2)}.
\end{align*}

\begin{figure*}[!t]
\centering
  \resizebox{13cm}{!}{\includegraphics{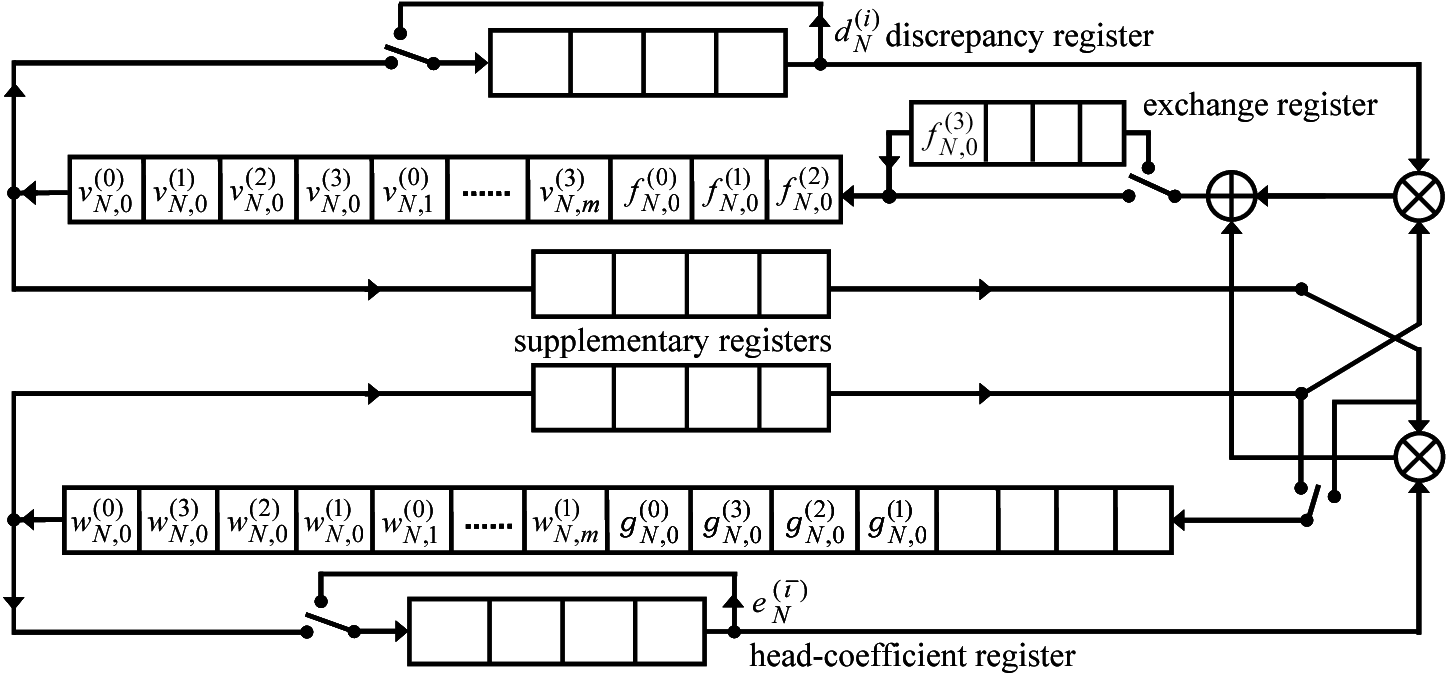}}
\caption{Serial inverse-free architecture for Hermitian codes, which is the 
closest to the RS-code error-locator ones.\label{serial inverse-free 
decoder}}
\end{figure*}
\begin{figure*}[!t]
\centering
  \resizebox{16cm}{!}{\includegraphics{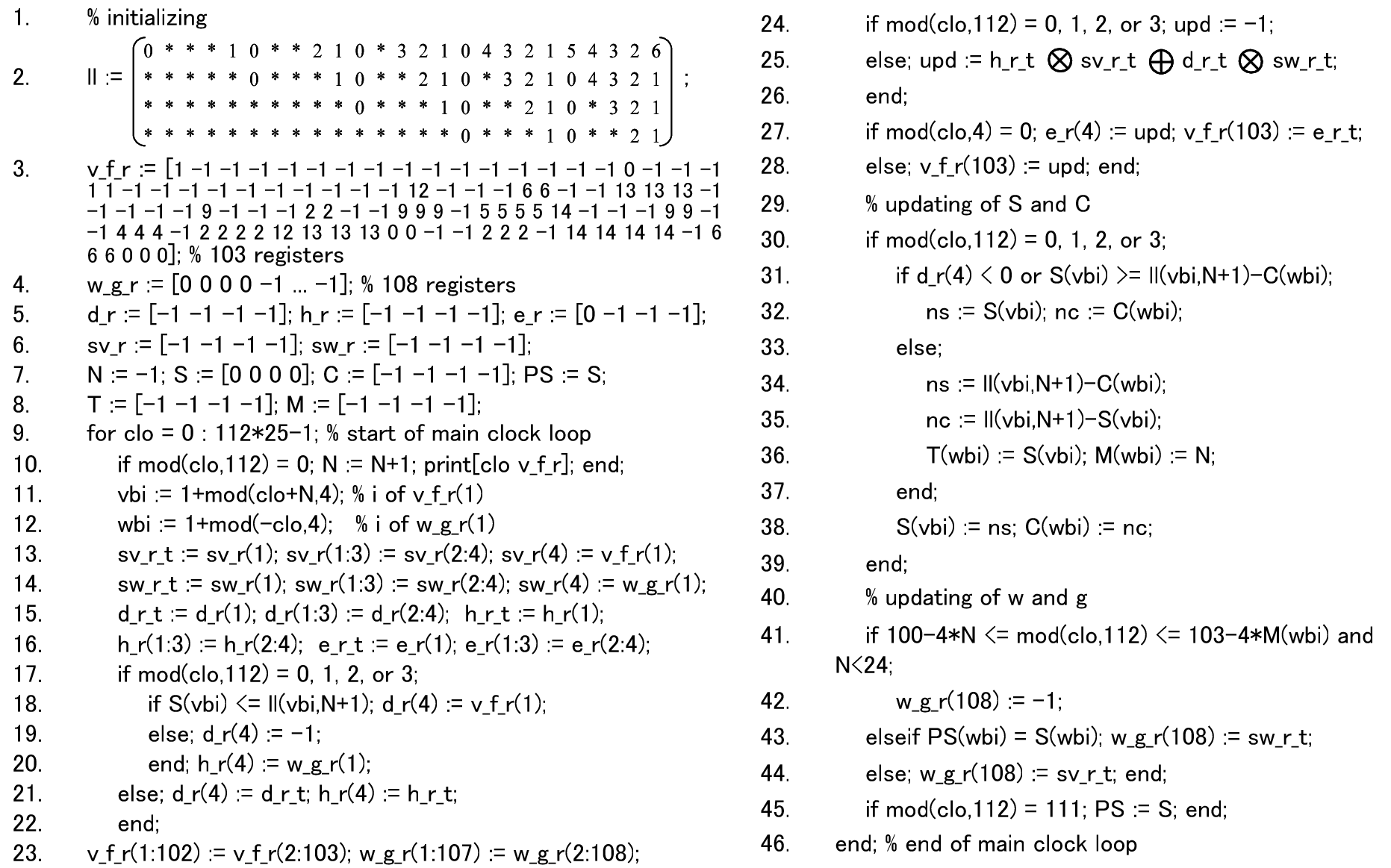}}
\caption{Program simulating the serial inverse-free architecture for $(64,45,14)$ Hermitian code over GF($2^4$) with five-error correction.\label{serial inverse-free program}}
\end{figure*}

Thus we have obtained the error-locator polynomials
\begin{align*}
F_{16}^{(0)}&=x^{3}+x^{2}+\alpha^{3}xy+\alpha^{2}x+\alpha,\\
F_{16}^{(1)}&=x^{2}y+\alpha x^{2}+\alpha^{6}xy+\alpha^{2}x+\alpha^{6},\\
F_{16}^{(2)}&=xy^{2}+\alpha^{2}x^{2}+xy+\alpha^{6}x+\alpha^{5},
\end{align*}
whose common zeros in the rational points decide $\mathcal{E}$,
and the auxiliary polynomials
\begin{align*}
G_{16}^{(0)}&=\alpha^{4}xy+\alpha^{6}x+\alpha^{6},\quad
G_{16}^{(1)}=0,\\
G_{16}^{(2)}&=\alpha^{4}x^{2}+\alpha^{6}x+\alpha^{4}.
\end{align*}
Then we obtain each error-value by O'Sullivan's formula
\cite{O'Sullivan error values}
$$e_{j}=\Bigg(\sum_{0\le i<a}
F_{m+1}^{(i)\,\prime}(P_{j})
G_{m+1}^{(i)}(P_{j})
\Bigg)^{-1}\;\mathrm{for}\;P_{j}\in\mathcal{E},$$
where $F_{m+1}^{(i)\,\prime}(z)$ is
the formal derivative of $F_{m+1}^{(i)}(z)$
with respect to $x$, e.g., $y^{\,\prime}=(x^{2}+y^{3})(xy^{2}+1)^{-1}$.
The divisions in \eqref{O'Sullivan's formula} are not required in this 
architecture since $F_{m+1,s}^{(i)}$ and $e_{m+1}^{(i)}$ have been 
normalized as $\alpha^{0}$.

The definite difference from the preceding one is that the serial 
architecture has a compact structure analogous to the RS-code case, with one 
inverse-calculator for the parallel BMS algorithm (not inverse-free).
In the next section, we will try to remove it from the serial architecture.

\section{Serial inverse-free architecture}\label{Serial inverse-free architecture}
We describe \textit{serial inverse-free architecture} \cite{SITA05}, which has the smallest 
circuit-scale we have ever obtained and is the last among the three kinds of proposed architectures.
In this section, we focus on Hermitian codes, that is, codes on Hermitian 
curves. These codes over ${\mathbb F}_{256}$ have the outstanding 
properties, and are ones of the most promising candidates for practical use.
For simplicity, here we simulate the architecture for a Hermitian code over 
$K:={\mathbb F}_{16}$.
The Hermitian curve defined by equation $y^{4}+y=x^{5}$ is one of 
C${}_{4}^{5}$ curves, and has 65 $K$-rational points equal to the 
Hasse--Weil upper bound with genus $6$.
Then codes on this curve can have code-length 64.

\begin{table*}[!t]
\caption{Values of registers in two shift-register sequences, discrepancy 
$d_{N}^{(i)}$, and $s_{N,1}^{(i)}$in the serial inverse-free 
architecture.\label{serial if register value}}
\centering
  \resizebox{17.5cm}{!}{\includegraphics{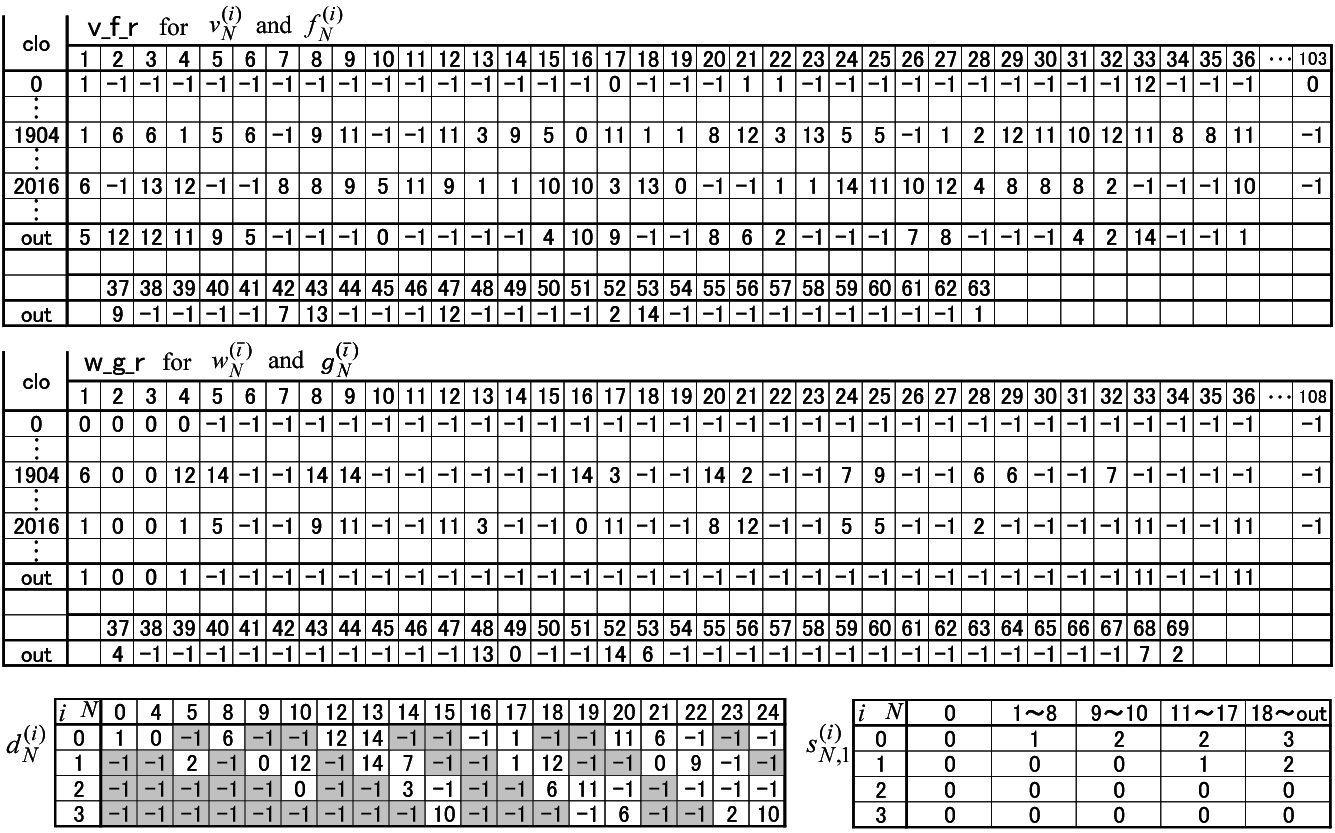}}
\end{table*}

As in the preceding two sections, we intend to correct generic errors in 
$\mathcal{C}(m+3)$ with $m:=2t+11$. The notations concerning $K$ are the 
same as in subsection \ref{Simulation}.
We demonstrate 5-error correction, and set the error-locations 
$\mathcal{E}:=\{(x,y)=(-1,0),(5,3),(9,8),(10,13),(12,2)\}$, and let error 
values be 11, 13, 2, 12, 9, respectively.

As shown in the model Fig. \ref{serial inverse-free decoder}, the serial 
inverse-free architecture also has the same single structure as that of RS 
codes.
Initially, the coefficients of $v_{N}^{(i)}$s and 
$f_{N}^{(i)}$s are arranged serially in the order $i=0,1,2,3$ in a sequence 
of shift-registers, and those of $w_{N}^{(\overline{\imath})}$s and 
$g_{N}^{(\overline{\imath})}$s are arranged in the order 
$\overline{\imath}=0,3,2,1$ in another.
This arrangement of coefficients is decided by the pair 
$(i,\,\overline{\imath})$ with 
$i+\overline{\imath}\equiv0\:(\mathrm{mod}\,4)$, and in general for other 
codes on C${}_{a}^{b}$ curves, one can also arrange them in a similar 
manner with $i+\overline{\imath}\equiv0\:(\mathrm{mod}\,a)$.
Then the exchange register changes the order $i=0,1,2,3$ of 
$\{v_{N}^{(i)},f_{N}^{(i)}\}$s at $N\equiv0\:(\mathrm{mod}\,4)$ into 
$i=1,2,3,0$ at $N\equiv1$, $\cdots$, $i=3,0,1,2$ at $N\equiv3$. In general,
for other codes on C${}_{a}^{b}$ curves, it changes the order of $i$ so as 
to keep $i+\overline{\imath}\equiv b^{-1}N\:(\mathrm{mod}\,a)$ as the 
definition of $\overline{\imath}$.

In the case of the serial inverse-free architecture, we require two other
sequences of $a$ shift-registers, \textit{supplementary registers}, as in 
Fig. \ref{serial inverse-free decoder}.
These do not appear in the algorithm but are due to technical
reasons in the architecture.
For example, we can see in Table \ref{serial if register value} that the 
values $s_{17,1}^{(0)}=2$ and $s_{17,1}^{(1)}=1$ are increased to 3 and 2 
at the same $N=18$.
For such cases, the supplementary registers hold the values of the head 
coefficients $v_{N,N}^{(i)}$ and $w_{N,N}^{(\overline{\imath})}$;
otherwise the value $w_{N,N}^{(\overline{\imath})}$ cannot be updated to 
$v_{N,N}^{(i)}$.

\begin{figure}[!t]
\centering
  \resizebox{7cm}{!}{\includegraphics{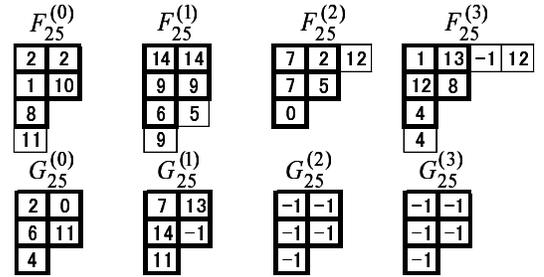}}
\caption{Output of the serial inverse-free architecture, where polynomials 
are depicted on $\Phi(4,9)$.\label{polynomials}}
\end{figure}

For the same reason as the previous ones, suitable register values should be 
set to zero at line {\sf41}, where the condition is derived by taking the supplementary registers into account as follows:
Since $\alpha^{0}=f_{N,0}^{(0)}$ is at the 101-th register in the initial 
values of {\sf v\underline{ }f\underline{ }r} as seen in line {\sf 3}, we 
claim that the head coefficient $g_{N,N-M^{(i)}}^{(i)}$ is located at the 
$(101-4M^{(i)})$-th register of {\sf w\underline{ }g\underline{ }r} if {\sf 
mod(clo,112)} $=i$. For example, if $N=18$, we can see from $s_{N,1}^{(i)}$ 
in Table \ref{serial if register value} that $M^{(0)}=M^{(1)}=17$. Then, in 
{\sf w\underline{ }g\underline{ }r(33)}, $g_{18,1}^{(0)}=\alpha^{11}$ is at 
{\sf clo} $=2016$,  and $g_{18,1}^{(1)}=\alpha^{11}$ is at {\sf clo} 
$=2019$.

Similarly as in section \ref{Serial architecture}, we note that the value in {\sf w\underline{ }g\underline{ }r(j)} at {\sf mod(clo,112)} $=i$ is the shifted value at {\sf mod(clo,112)} $=i+j-1+4$, where ``$+4$'' is caused by the supplementary four shift-registers.
Moreover, since each $N+1-M^{(i)}$ value of {\sf w\underline{ }g\underline{ 
}r(j)} for $j=97-4N$, $97-4N+4$, $\cdots$, $97-4M^{(i)}$ must be $-1$ at 
{\sf mod(clo,112)} $=i$ in each $w_{N}^{(i)}$,
we obtain the upper and lower conditions of {\sf w\underline{ }g\underline{ 
}r(108)} $:=-1$ at line {\sf41} as the union of
\begin{align*}
i=0\;\Rightarrow\;j&=100-4N,\,\cdots,\,100-4M^{(0)},\\
&\vdots\\
i=3\;\Rightarrow\;j&=103-4N,\,\cdots,\,103-4M^{(3)}.
\end{align*}

Thus, the Gr\"obner basis of ideal $I(\mathcal{E})$ and the auxiliary 
polynomials have been obtained as in Fig. \ref{polynomials}, e.g.,
$$
F_{25}^{(0)}=\alpha^{11}x^{3}+\alpha^{10}xy+\alpha^{8}x^{2}+\alpha^{2}y
+\alpha x+\alpha^{2},
$$
and obtain each error-value by O'Sullivan's formula
\eqref{O'Sullivan's formula}.

In this manner, we have constructed the smallest-scale architecture, which uses the supplementary registers differently from the others.
In our example, the total number of shift-registers for polynomials is 215, 
while for the supplementary registers, it is 8, i.e., 3.7\%.
Furthermore, this percentage is decreased for larger $t$, and approximately $1/m$, as seen in the next section;
we have, e.g., $m=2t+239$ for the other Hermitian codes over ${\mathbb 
F}_{256}$. Hence we can say that $2a$ shift-registers for the supplementary 
registers are reasonably small in the whole architecture.

\section{Performance estimation}\label{Performance estimation}

In this section, we estimate the numbers of multipliers, calculators for 
inverse, and registers, and the total running time.
Although the estimation at Section IX in \cite{IEEE05} was done with respect 
to the upper bound $\lambda=t+2g-1+a$ of $o(s_{N}^{(i)})$s, it is now 
convenient to estimate with respect to $m=2t+2g-1$ of the code 
$\mathcal{C}(m)$ since we consider architectures without the determination 
of unknown-syndrome values.

We quote the result of the systolic array in \cite{IEEE05}; the numbers of 
multipliers and calculators for inverse are $2am$ and $am/2$, respectively, 
as seen at the upper part of Fig.4 in [p.3866,\ref{IEEE05}].
The number of registers and the total running time are $(4m+9)a/2$ and 
$m+1$, respectively.

The K\"otter's architecture \cite{K"otter} has $3a$ multipliers, $a$ 
calculators for inverse, and $a(4\lambda+5)$ registers, where 
$\lambda=(m+1)/2-1+a$ since we restrict correctable errors to the generic 
errors.
The total running time takes $2(\lambda+1)(m+1)=(m+3)(m+1)$.

The serial architecture and the serial inverse-free architecture have two multipliers, and the inverse-free architecture has $a$ times two multipliers.
There is one calculator for inverse only in the serial architecture.
The number of registers for these three architectures is equal to $2a$ times
$m+2$, which consists of the number of syndromes including the gaps plus one for the initial value of $f_{N}^{(i)}$; we ignore the contribution 
of the discrepancy, exchange, and supplementary registers since these are at 
most a few multiples of $a$ and disappear in the order of $m$.
The total running time for the inverse-free architecture agrees with $m+1$ 
times the number of registers in the sequence for $w_{N}^{(i)}$ and 
$g_{N}^{(i)}$, which is equal to $(m+1)(m+2)$.
Those for the other two agree with $a(m+1)(m+2)$.

We summarize these results in Table \ref{performance}, where we denote only 
the terms of the highest orders for $m$ in the estimations.
In addition, there is an architecture between K\"otter's and Inverse-free 
that employs the parallel BMS algorithm (not inverse-free); we call 
this temporarily \textit{parallel-BMS architecture} and add it to the table.
For example, in the case of Hermitian codes over $2^{8}$-element finite 
field, $a$ and $m$ is equal to 16 and $2t+239$, respectively.
Since the numbers of registers in all architectures have an unchanged order $2am$ in Table \ref{performance}, we can see that these architectures have 
optimized their space complexity.

Then we can see in Table \ref{performance} that $a$ multipliers have been 
reduced from K\"otter's to Parallel-BMS, and that $a$ inverse-calculators 
have been reduced from Parallel-BMS to Inverse-free. Both contribute to 
the reduction of computational complexity. It is noticed that the latter 
reduction has been accompanied in $\mathcal{C}(m+a-1)$ by the slight decrease $\left\lfloor\frac{a-1}{2}\right\rfloor$ of correctable errors that is assignable to error-detection.
On the other hand, two types of serial architectures have the constant 
numbers of finite-field calculators, and their running time takes $a$ times 
longer than that of non-serial types.
Thus our serializing method has provided a preferred trade-off between 
calculators and delay.

\begin{table}[!t]
\caption{Performance of various architectures.\label{performance}}
\centering
  \resizebox{7.5cm}{!}{\includegraphics{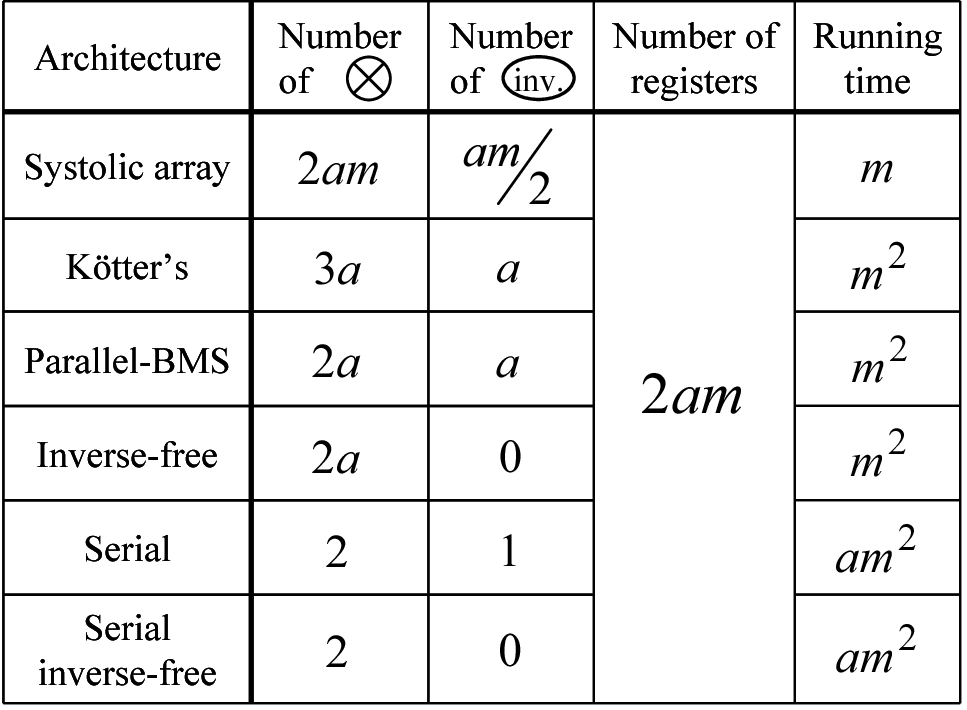}}
\end{table}

\section{Conclusions}\label{Conclusions}

In this paper, we have proposed the inverse-free parallel BMS algorithm for 
error-location in decoding algebraic-geometric codes.
Thus we have improved decoding bound $t\le\left\lfloor(d_\mathrm{G}-g-1)/2\right\rfloor$ in \cite{Justesen} based on linear system without the determination of unknown syndromes for AG codes, to $t\le\left\lfloor(d_\mathrm{FG}-a)/2\right\rfloor$ for generic errors, where, e.g., $g=120$ and $a=16$ for Hermitian codes over $\mathbb{F}_{2^8}$.
Moreover, we have constructed three kinds of error-locator architectures 
using our algorithm.
These architectures were not implemented until the determination 
procedure of unknown syndromes was removed from the error-location algorithm.
Our novel algorithm and architectures have a wide range of applications to 
Gr\"obner-basis schemes in various algebraic-coding situations, such as Sudan algorithm \cite{Sudan}, Guruswami--Sudan algorithm \cite{Guruswami-Sudan}, Koetter--Vardy algorithm \cite{Koetter-Vardy}, and encoding of algebraic codes \cite{ISIT07}.

We have aimed to construct our architectures with only shift-registers, 
switches, and finite-field calculators.
The composition of shift-registers is superior to that of RAMs 
(random-access memories) in decoding speed, and moreover, our approach is 
useful for revealing their regularity.

We can conclude that the error-locator architectures correcting generic 
errors have been completed by the whole from systolic array (max. 
parallelism) to serial inverse-free ones (min. parallelism).
These architectures enable us to fit the decoder of the codes to 
various sizes and speeds in many applications.
It may also be concluded that our methodology, which is the direct decoding 
from only the received syndromes, correctly generalizes the RS-code case.

\appendices
\section{Proof that $V(u,A)$ is an ideal}\label{V(u,A)}
We first note that, by \eqref{discrepancy} and the following lemma,
\begin{align}
f\in&\,V(u,A)\Leftrightarrow df_{l}=0\;\mathrm{for}\;l\in\Phi(a,A)
\nonumber\\
&\Leftrightarrow\sum_{n\in\Phi(a,s)}f_{n}u_{n+h}=0
\;\mathrm{for}\;h\in\Phi(a,A-o(s)).\label{type}
\end{align}
\begin{Lemma}\label{surprising}
We have
$\{l^{(s_{2})}-s\,|\,l\in\Phi(a,A),\,l^{(s_{2})}\ge s\}
=\Phi(a,A-o(s))$.\hfill$\Box$
\end{Lemma}
\textit{Proof.}\quad
Obviously
$\{l^{(s_{2})}-s\,|\,l\in\Phi(a,A),\,l^{(s_{2})}\ge s\}$ equals
$$\{l-s\,|\,l\in\Phi^{(s_{2})}(a,A),\,l\ge s\}=\Phi(a,A-o(s)),$$
where the last equality follows from correspondence 
$l-s=:h\in\Phi(a,A-o(s))$.\hfill$\Box$

For simplicity, we denote $P_{j}$ and $e_{j}$ as 
$P_{\gamma_{j}}\in\mathcal{E}$ and the error-value $e_{\gamma_{j}}$ without 
loss of generality.
Then we convert the sum $\sum f_{n}u_{n+h}$ in \eqref{type} as
\begin{align}
\sum_{n\in\Phi(a,s)}\hspace{-2mm}f_{n}\sum_{j=1}^{t}
e_{j}z^{n+h}(P_{j})
&=\sum_{j=1}^{t}e_{j}z^{h}(P_{j})\hspace{-2mm}
\sum_{n\in\Phi(a,s)}f_{n}z^{n}(P_{j})\nonumber\\
&=\sum_{j=1}^{t}e_{j}z^{h}\label{convert}
(P_{j})f(P_{j}).
\end{align}

\begin{Proposition}\label{ideal}
For all $A\in{\mathbb Z}_{0}$, the set $V(u,A)\subset K[\mathcal{X}]$ is a 
polynomial ideal.\hfill$\Box$
\end{Proposition}

\textit{Proof.}\quad
Suppose that $f$ and $g\in V(u,A)$ with $s:=\deg(f)$ and $t:=\deg(g)$. Then 
we show that $f+g$ and $z^{h}f\in V(u,A)$. Note that, by \eqref{convert},
\begin{align*}
d(f+g)_{l}&=\sum_{j=1}^{t}e_{j}(f+g)(P_{j})z^{l^{(s_{2}+t_{2})}-s-t}(P_{j})\\
&=\sum e_{j}f(P_{j})z^{l^{(s_{2}+t_{2})}-s-t}(P_{j})\\
&\qquad\qquad+\sum e_{j}g(P_{j})z^{l^{(s_{2}+t_{2})}-s-t}(P_{j}),
\end{align*}
and the last two sums are zero from the assumption and 
$\{l^{(s_{2}+t_{2})}-s-t\}=\Phi(a,A-o(s)-o(t))\subset\Phi(a,A-o(s))$, 
$\Phi(a,A-o(t))$ by Lemma \ref{surprising}.
For $z^{h}f$, note that
\begin{align*}
d(z^{h}f)_{l}&=\sum e_{j}(z^{h}f)(P_{j})z^{l^{(s_{2}+h_{2})}-s-h}(P_{j})\\
&=\sum e_{j}f(P_{j})z^{l^{(s_{2}+h_{2})}-s}(P_{j}),
\end{align*}
and $\{l^{(s_{2}+h_{2})}-s\}=\Phi(a,A-o(s)-o(h))+h$ by Lemma 
\ref{surprising}.
Although $\Phi(a,A-o(s)-o(h))+h\not\subset\Phi(a,A-o(s))$ in general, the 
monomial $z^{l^{(s_{2}+h_{2})}-s}$ is represented as the linear combination 
of elements in $\{z^{l}\,|\,l\in\Phi(a,A-o(s))\}$. Then we obtain 
$d(z^{h}f)_{l}=0$ from the assumption, which completes the 
proof.\hfill$\Box$

\section{Proof of $V(u,B)=I(\mathcal{E})$}\label{accord}

This follows from the next Corollary and Lemma \ref{bound}.

\begin{Proposition}\label{determinant}
Let $f\in K[\mathcal{X}]$ be satisfying
$$
\sum_{h\in\Phi(a,s)}f_{h}u_{h+l_{j}}=0\;\mathrm{for}\;
l_{j}\in\Phi(a)\;\mathrm{with}\;j=1,\cdots,t
$$
and $\det\left(\left[z^{l_{j}}(P_{j'})\right]\right)\not=0$.
Then $f\in I(\mathcal{E})$ holds.\hfill$\Box$
\end{Proposition}

\textit{Proof.}\quad
Since $\sum_{h\in\Phi(a,s)}f_{h}u_{h+l}$ is converted as 
\eqref{convert}.\hfill$\Box$
\vspace{1em}

Using Riemann--Roch Theorem, we see that
the map
$$L((t+2g-1)P_{\infty})\rightarrow{\mathbb F}_{q}^{t}\quad
(f\mapsto \left[f(P_{1}),\cdots,f(P_{t})\right])$$ is surjective.
Hence there are linearly independent $t$ vectors of the form
$\left[z^{l}(P_{1}),\cdots,z^{l}(P_{t})\right]$
for $l\in\Phi(a,t+2g-1)$,
and we obtain the following sufficient condition for all errors.

\begin{Corollary}\label{Riemann-Roch}
Let $f\in K[\mathcal{X}]$ be satisfying
$\displaystyle\sum_{h\in\Phi(a,s)}f_{h}u_{h+l}=0$ for all
$l\in\Phi(a,t+2g-1)$.
Then $f\in I(\mathcal{E})$ holds.\hfill$\Box$
\end{Corollary}

\begin{Lemma}\label{bound}
We can choose a Gr\"obner basis $\{f^{(i)}\}_{0\le i<a}$ of $I(\mathcal{E})$ 
as $o(f^{(i)})\le t+2g-1+a$ for all $i$.\hfill$\Box$
\end{Lemma}

\textit{Proof.}
First, we notice that an element $f^{(i)}$ of Gr\"obner basis may be 
determined uniquely by
\begin{equation}\label{Weierstrass}
o(f^{(i)})=\min_{f\in I(\mathcal{E})}
\big\{o(f)\,\big|\,o(f)\equiv i\:\mathrm{mod}\:a\big\}.
\end{equation}
Let $n_{i}$ be one of
$\{t+2g,\,t+2g+1,\,\cdots,t+2g-1+a\}$
satisfying $n_{i}\equiv\,i\,\mathrm{mod}\,a$.
We temporarily denote as $\ell(D):=\dim L(D)$,
where $L(D):=\{f\in K[\mathcal{X}]\,|\,
\mathrm{divisor}(f)+D\;\mathrm{is}\;\mathrm{positive}\}\cup\{0\}$
for a divisor $D$.
Since we have
\begin{align*}
\ell\big((t+2g-1)P_{\infty}-E\big)&=g,\\
\ell\big((t+2g)P_{\infty}-E\big)&=g+1,\\
&\vdots\\
\ell\big((t+2g-1+a)P_{\infty}-E\big)&=g+a,
\end{align*}
where $E:=\sum_{j=1}^{t}P_{j}$,
there is $f\in I(\mathcal{E})$ satisfying
$o(f)=n_{i}$.
Then $o(f^{(i)})\le n_{i}$ is obtained by \eqref{Weierstrass},
and
$\max\{o(f^{(i)})\,|\,0\le i<a\}
\le\max\{n_{i}\,|\,0\le i<a\}=t+2g-1+a
$
leads Lemma \ref{bound}.\hfill$\Box$
\vspace{1em}

\begin{Proposition}\label{ideal minimal set}
$B\ge 2t+4g-2+a\;\Rightarrow\;V(u,B)=I(\mathcal{E})$\hfill$\Box$
\end{Proposition}

\textit{Proof.}
If $f\in K[\mathcal{X}]$ and $s:=\deg(f)\le l^{(s_{2})}$, then $df_{l}$ is 
converted similarly as \eqref{convert} to
$$
df_{l}:=\sum e_{i}f(P_{i})z^{l^{(s_{2})}-s}(P_{i}).
$$
Hence, if $f(P_{1})=\cdots=f(P_{t})=0$, then we have $df_{l}=0$, and thus 
$I(\mathcal{E})\subset V(u,B)$ is obvious.
To prove $\supset$, let $\{f^{(i)}_{B+1}\}_{0\le i<a}$ be a Gr\"obner basis 
of $V(u,B)$, where ``$B+1$'' is for consistency in the previous 
notation. Since $I(\mathcal{E})\subset V(u,B)$, we can choose it as 
$o(f^{(i)}_{B+1})\le t+2g-1+a$ from Lemma \ref{bound} and its proof. Now we 
suppose that 
$d(f^{(i)}_{B+1})_{l}=\sum_{h}f^{(i)}_{B+1,h}u_{h+l^{(i)}-s_{B+1}^{(i)}}=0$ 
for all $l\in\Phi(a,B)$ with $l^{(i)}\ge s_{B+1}^{(i)}$. Then we have, by 
Lemma \ref{surprising},
$\{l^{(i)}-s_{B+1}^{(i)}\}=\Phi(a,B-o(s_{B+1}^{(i)}))\subset\Phi(a,t+2g-1)$. 
Thus we see that the inverse inclusion follows from Corollary 
\ref{Riemann-Roch} of Proposition \ref{determinant}.\hfill$\Box$

\section{Generic case}\label{Generic case}

Let $m_{t}:=\min\left\{m\in{\mathbb Z}_{0}\left|\,\dim 
L(mP_{\infty})=t\right.\right\}$; recall that $\dim L(mP_{\infty})$ is equal 
to the number of $l\in\Phi(a,m)$. If $t>g$, then we have $m_{t}=t+g-1$ since 
$\dim L((t+g-1)P_{\infty})=t$ and $\dim L((t+g-2)P_{\infty})=t-1$. However, 
for $t\le g$, we have for example $m_{6}=10<t+g-1$ for Hermitian curve 
$y^{4}+y=x^{5}$ over $\mathbb{F}_{2^{4}}$.
We define that $t$-error position $\mathcal{E}$ is 
\textit{generic} if $\det\left(\left[z^{l_{j}}(P_{j'})\right]\right)\not=0$
for $P_{j'}\in\mathcal{E}$ and $l_{j}\in\Phi(a,m_{t})$.
If $\mathcal{E}$ is generic, we obtain a Gr\"obner basis
$\left\{f^{(i)}=z^{s^{(i)}}-\sum_{l_{j}\in\Phi(a,m_{t})}f_{l_{j}}^{(i)}z^{l_{j}}\right\}$ of $I(\mathcal{E})$ by solving
$$
\left[\begin{array}{ccc}
z^{l_{1}}(P_{1})&\cdots&z^{l_{t}}(P_{1})\\
\vdots&\empty&\vdots\\
z^{l_{1}}(P_{t})&\cdots&z^{l_{t}}(P_{t})\end{array}
\right]
\left[\begin{array}{c}
f^{(i)}_{l_{1}}\\
\vdots\\
f^{(i)}_{l_{t}}\end{array}
\right]
=
\left[\begin{array}{c}
z^{s^{(i)}}(P_{1})\\
\vdots\\
z^{s^{(i)}}(P_{t})\end{array}
\right]
$$
with $s^{(i)}\in\Phi(a,m_{t+i+1})\backslash\Phi(a,m_{t+i})$.
Then Lemma \ref{bound} is improved to $o(f^{(i)})\le t+g-1+a$ for generic 
$\mathcal{E}$.

Conversely, if $\det\left(\left[z^{l_{j}}(P_{j'})\right]\right)=0$, then the equation from the linear dependency gives $f\in I(\mathcal{E})$ with $\mathrm{deg}(f)\in\Phi(a,m_{t})$.
Thus we see that $\mathcal{E}$ is generic if and only if the delta set $\{l\in\Phi(a)\,|\,l\le s^{(l_{2})}\}$ (footprint in \cite{ISITA04}) agrees with $\Phi(a,m_{t})$.
Namely, our definition of generic is equivalent to the definition of generic in \cite{generic} and that of ``independent'' in \cite{independent}.

\begin{Proposition}\label{proof}
Suppose that $\mathcal{E}$ is generic.\\
If $f\in V(u,m_{t}+o(f))$, then we have $f\in I(\mathcal{E})$.
In particular, $V(u,m+a-1)=I(\mathcal{E})$ with $m=2t+2g-1$.
\hfill$\Box$
\end{Proposition}

\textit{Proof.}\quad
Since $\{l^{(s_{2})}-s\,|\,l\in\Phi(a,m_{t}+o(f)),\,l^{(s_{2})}\ge s\}$ 
agrees with $\Phi(a,m_{t})$ by Lemma \ref{surprising}, it follows from 
Proposition \ref{determinant}.\hfill$\Box$

\section{Proof of Theorem \ref{induction}}\label{Proof of Theorem}
Theorem \ref{induction} is proved by the following three lemmas.

\begin{Lemma}\label{minimalityfollows}
Suppose that $G(z)\in V(u,M-1)$, $dG_{k}\not=0$,
and $t\le k$ with $t=\mathrm{deg}(G)$,
$k\in\Phi^{(t_{2})}(a,M)$, and $o(k)=M$.
Moreover, suppose that $F(z)\in V(u,M)$ and $dF_{s}\not=0$
with $s=\mathrm{deg}(F)$.
Then, at least one condition of
$s_{1}\ge k_{1}-t_{1}+1$ and $s_{2}\not=k_{2}-t_{2}$ holds.\hfill$\Box$
\end{Lemma}

\textit{Proof.}
We suppose that $s_{1}\le k_{1}-t_{1}$ and $s_{2}=k_{2}-t_{2}$.
Since $G\in V(u,M-1)$ and $F\in V(u,M)$, we have
\begin{align*}
&-\hspace{-4mm}\sum_{n\in\Phi(a,t)\backslash\{t\}}
\hspace{-3mm}G_{n}u_{n+l-t}=G_{t}u_{l}
\;\,\mathrm{for}\;\,
l\in\Phi^{(t_{2})}(a,M-1),\;t\le l,\\
&-\hspace{-2mm}\sum_{r\in\Phi(a,s)\backslash\{s\}}
\hspace{-3mm}F_{r}u_{r+l-s}=F_{s}u_{l}
\;\,\mathrm{for}\;\,
l\in\Phi^{(s_{2})}(a,M),\;s\le l.
\end{align*}
Since $n_{2}+k_{2}-t_{2}\le a-1+s_{2}$ and $n+k-t\ge n+s\ge s$
for $n\in\Phi(a,t)$,
we have $n+k-t\in\Phi^{(s_{2})}(a,M)$ and $s\le n+k-t$
for $n\in\Phi(a,t)$,
and moreover,
\begin{align*}
&\quad-\sum_{n\in\Phi(a,t)\backslash\{t\}}G_{n}u_{n+k-t}\\
&=\sum_{n\in\Phi(a,t)\backslash\{t\}}\hspace{-2mm}G_{n}
\left\{\frac{1}{F_{s}}\sum_{r\in\Phi(a,s)\backslash\{s\}}
\hspace{-2mm}F_{r}u_{r+(n+k-t)-s}\right\}\\
&=\frac{1}{F_{s}}\sum_{r\in\Phi(a,s)\backslash\{s\}}
\hspace{-1mm}F_{r}\hspace{-1mm}
\sum_{n\in\Phi(a,t)\backslash\{t\}}\hspace{-2mm}G_{n}u_{n+(r+k-s)-t}\\
&=-\frac{G_{t}}{F_{s}}\sum_{r\in\Phi(a,s)\backslash\{s\}}F_{r}u_{r+k-s},
\end{align*}
where the last equality follows from $r+k-s\in\Phi^{(t_{2})}(a,M-1)$ and
$t\le r+k-s$ for $r\in\Phi(a,s)\backslash\{s\}$
since $r_{2}+k_{2}-s_{2}\le a-1+t_{2}$ and $r+k-s\ge r+t\ge t$
for $r\in\Phi(a,s)$,
and the last sum agrees with $G_{t}u_{k}$
since $s_{2}\le k_{2}=s_{2}+t_{2}\le s_{2}+a-1$
and $k\in\Phi^{(s_{2})}(a,M)$.
This contradicts $dG_{k}\not=0$.\hfill$\Box$

\begin{Lemma}\label{s=c+1}
We have $s_{N,1}^{(i)}=c_{N,1}^{(i)}+1$.\hfill$\Box$
\end{Lemma}

\textit{Proof.}
We prove it by induction.
The case of $N=0$ follows from the initializing.
Assuming $s_{N,1}^{(i)}=c_{N,1}^{(i)}+1$ for all $i$,
we prove $s_{N+1,1}^{(i)}=c_{N+1,1}^{(i)}+1$.
We may assume that there is
$l^{(i)}=l^{(\overline{\imath})}$.
It follows that
$s_{N,1}^{(i)}\ge l_{1}^{(i)}-c_{N,1}^{(\overline{\imath})}
\Leftrightarrow
s_{N,1}^{(\overline{\imath})}\ge l_{1}^{(\overline{\imath})}-c_{N,1}^{(i)}$.
Thus we may assume that
$s_{N,1}^{(i)}<l_{1}^{(i)}-c_{N,1}^{(\overline{\imath})}$,
$s_{N,1}^{(\overline{\imath})}<l_{1}^{(\overline{\imath})}-c_{N,1}^{(i)}$,
and $d_{N}^{(i)}\not=0$ without loss of generality.
If $d_{N}^{(\overline{\imath})}=0$,
then it contradicts Lemma \ref{minimalityfollows}
since $F_{N}^{(i)}\in V(u,N-1)$, $F_{N}^{(\overline{\imath})}\in V(u,N)$,
$s_{N,1}^{(\overline{\imath})}\le l_{1}^{(i)}-s_{N,1}^{(i)}$,
and $\overline{\imath}=l_{2}^{(i)}-i$.
Thus, we obtain $d_{N}^{(\overline{\imath})}\not=0$ and
$s_{N+1,1}^{(i)}-c_{N+1,1}^{(i)}
=s_{N,1}^{(\overline{\imath})}-c_{N,1}^{(\overline{\imath})}$.
\hfill$\Box$

\begin{Lemma}\label{Berlekamptransform}
Let $F(z)\in V(u,N-1)$, $s\le l$
with $s=\mathrm{deg}(F)$ for $l\in\Phi^{(s_{2})}(a,B)$,
and let $G(z)\in V(u,M-1)$, $t\le k$
with $t=\mathrm{deg}(G)$ for $k\in\Phi^{(t_{2})}(a,B)$.
Suppose that $dG_{k}\not=0$,
$M=o(k)<N=o(l)$ and $k_{2}-t_{2}=l_{2}-s_{2}$.
Then we have
$$
H(z):=
dG_{k}z^{r-s}F
-dF_{l}z^{r-l+k-t}G\;\:\in V(u,N),
$$
and $\mathrm{deg}(H)=r$,
where $r:=s$ if $dF_{l}=0$, and
$r:=\left(\max\{s_{1},l_{1}-k_{1}+t_{1}\},s_{2}\right)$ otherwise.
\hfill$\Box$
\end{Lemma}

\textit{Proof.}
Since $r_{2}=s_{2}$ and
\begin{align}
&\quad o\left(z^{r-s}F\right)-o\left(z^{r-l+k-t}G\right)
\nonumber\\
&=r_{1}a+s_{2}b-(r_{1}-l_{1}+k_{1})a-t_{2}b\label{555}\\
&=o(l)-o(k)>0,\nonumber
\end{align}
we obtain $\mathrm{deg}(H)=r$.
Next, since $F\in V(u,N-1)$ and $G\in V(u,M-1)$, we have
\begin{align*}
\sum_{n\in\Phi(a,s)}
\hspace{-2mm}F_{n}u_{n+p-s}&=
\left\{\begin{array}{cl}
  0 & p\in\Phi^{(s_{2})}(a,N-1),\,s\le p\\
dF_{l} & p=l,
        \end{array}\right.\\
\sum_{n\in\Phi(a,t)}
\hspace{-2mm}G_{n}u_{n+p-t}&=
\left\{\begin{array}{cl}
  0 & p\in\Phi^{(t_{2})}(a,M-1),\,t\le p\\
dG_{k} & p=k.
        \end{array}\right.
\end{align*}
We may assume that $dF_{l}\not=0$.
If $p\in\Phi^{(s_{2})}(a,N-1)$ and $r\le p$,
then we have
$p-l+k\in\Phi^{(t_{2})}(a,M-1)$ and $t\le p-l+k$
from $l-k+t\le r$, and moreover,
\begin{align*}
&\qquad\sum_{n\in\Phi(a,r)}H_{n}u_{n+p-r}\\
&=dG_{k}\hspace{-4mm}\sum_{n\in\Phi(a,s)}\hspace{-3mm}F_{n}u_{n+(r-s)+p-r}
-dF_{l}\hspace{-3mm}\sum_{n\in\Phi(a,t)}\hspace{-3mm}
G_{n}u_{n+(r-l+k-t)+p-r}\\
&=dG_{k}\hspace{-2mm}\sum_{n\in\Phi(a,s)}\hspace{-2mm}F_{n}u_{n+p-s}
-dF_{l}\hspace{-2mm}\sum_{n\in\Phi(a,t)}\hspace{-2mm}G_{n}u_{n+(p-l+k)-t}\\
&=\left\{\begin{array}{cl}
  0 & \hspace{-1cm}p\in\Phi^{(s_{2})}(a,N-1),\,r\le p\\
dG_{k}\cdot dF_{l}-dF_{l}\cdot dG_{k}=0 & p=l.
        \end{array}\right.
\;\Box
\end{align*}

\textit{Proof of Theorem \ref{induction}.}
If $d_{N}^{(i)}\not=0$ and $G_{N}^{(\overline{\imath})}=0$,
then $s_{N+1,1}^{(i)}:=l_{1}^{(i)}+1$ and
$F_{N+1}^{(i)}:=x^{l_{1}^{(i)}+1}F_{N}^{(i)}$.
Thus $d_{N+1}^{(i)}=0$ and $\mathrm{deg}(F_{N+1}^{(i)})=s_{N+1}^{(i)}$ hold.
Supposing that $G_{N}^{(\overline{\imath})}\not=0$,
let $M<N$ be satisfying
$G_{N}^{(\overline{\imath})}:=\left(d_{M}^{(j)}\right)^{-1}F_{M}^{(j)}$,
$o(k^{(j)})=M$, and $\overline{\imath}=k_{2}^{(j)}-j$,
then we have $c_{N}^{(\overline{\imath})}=k^{(j)}-s_{M}^{(j)}$.
Thus the theorem except for \eqref{inequality} and \eqref{minimality}
follows from Lemma \ref{Berlekamptransform}.
We prove \eqref{minimality} by induction.
The case of $N=0$ in \eqref{minimality} holds by the definition.
Supposing that the equality is true for $s_{N,1}^{(i)}$,
we prove it for $s_{N+1,1}^{(i)}$.
Let $\varsigma_{N,1}^{(i)}$ be the minimum of $\zeta_{N,1}^{(i)}$
in \eqref{minimality}.
If (P),
then $s_{N,1}^{(i)}=\varsigma_{N,1}^{(i)}
\le\varsigma_{N+1,1}^{(i)}\le s_{N+1,1}^{(i)}=s_{N,1}^{(i)}$,
thus $\varsigma_{N+1,1}^{(i)}=s_{N+1,1}^{(i)}$ holds.
If $d_{N}^{(i)}\not=0$ and
$s_{N}^{(i)}>l^{(i)}-c_{N}^{(\overline{\imath})}$,
then we have $d_{N}^{(\overline{\imath})}\not=0$
as in the proof of Lemma \ref{s=c+1}
and $\varsigma_{N+1,1}^{(i)}\le s_{N+1,1}^{(i)}
=l_{1}^{(i)}-s_{N,1}^{(\overline{\imath})}+1$,
which is actually the equation $\varsigma_{N+1,1}^{(i)}=s_{N+1,1}^{(i)}$
by Lemma \ref{minimalityfollows} for
$F_{N}^{(\overline{\imath})}\in V(u,N-1)$ and $F\in V(u,N)$
satisfying $\mathrm{deg}(F)=(\varsigma_{N+1,1}^{(i)},i)$.
Finally, as for \eqref{inequality},
if we suppose $s_{N,1}^{(i)}<s_{N,1}^{(j)}$ with $i<j$,
then we have $y^{j-i}F_{N}^{(i)}\in V(u,N-1)$
and $\mathrm{deg}(y^{j-i}F_{N}^{(i)})=(s_{N,1}^{(i)},j)$,
which contradict the minimality of $s_{N,1}^{(j)}$.\hfill$\Box$

Thus we have proved the theorem for an algorithm that is not a parallel 
version, i.e., the algorithm with direct calculation of $d_{N}^{(i)}$ by 
\eqref{discrepancy} without $v_{N}^{(i)}$ and $w_{N}^{(i)}$.
To prove our parallel inverse-free BMS algorithm described in Section 
\ref{Inverse-free BMS algorithm}, we have to show further that $d_{N}^{(i)}$ 
is obtained by the coefficient of $v_{N}^{(i)}$; we omit this procedure and 
refer to similar cases \cite{Matsui-Sakata-Kurihara}\cite{IEEE05} of
ordinary parallel BMS algorithm.

\end{document}